\pdfoutput=1
\documentclass[cernpreprint, atlasdraft=false, texlive=2020, UKenglish, texmf]{atlasdoc}
\usepackage[full]{atlaspackage}
 
\usepackage{atlasbiblatex}
 
\usepackage[BSM,hepparticle,process,jetetmiss,texmf]{atlasphysics}
\usepackage{PMG}
 
\graphicspath{{logos/}{figures/}}
 
\usepackage{ANA-HDBS-2018-07-PAPER-defs}
 
\usepackage{booktabs}
\usepackage{siunitx}
\usepackage{tablefootnote}
\newcommand{\hsp}{\hspace*{0.3cm}}
 
\usepackage{multirow}
 
\usepackage{mathrsfs}
 
\AtlasTitle{Search for exotic decays of the Higgs boson into $b\bar{b}$ and missing transverse momentum in $pp$ collisions at $\sqrt{s} = 13$ TeV with the ATLAS detector }

\AtlasAbstract{
A search for the exotic decay of the Higgs boson ($H$) into a $b\bar{b}$
resonance plus missing transverse momentum is
described.    The search is performed with the ATLAS detector at the
Large Hadron Collider
using 139 $\mathrm{fb}^{-1}$ of $pp$ collisions at $\sqrt{s} = 13$ TeV.  The search
targets events from $ZH$ production in an NMSSM scenario where
$H \rightarrow \tilde{\chi}^{0}_{2}\tilde{\chi}^{0}_{1}$, with
$\tilde{\chi}^{0}_{2} \rightarrow {a} \tilde{\chi}^{0}_{1}$, where $a$ is a light
pseudoscalar Higgs boson and $\tilde{\chi}^{0}_{1,2}$ are the two
lightest neutralinos.  The decay of the $a$ boson into a pair of
$b$-quarks results in a peak in the dijet invariant mass distribution.  The
final-state signature consists of two leptons, two or more jets, at least one of which is identified as originating from a $b$-quark,
and missing transverse momentum.
Observations are consistent with Standard Model expectations and
upper limits are set on the product of cross section
times branching ratio
for a three-dimensional scan of the masses of the
$\tilde{\chi}^{0}_{2}$, $\tilde{\chi}^{0}_{1}$  and $a$
boson.
}
 
\author{The ATLAS Collaboration}
 
\AtlasDate{\today}

\AtlasJournal{JHEP}
 
\PreprintIdNumber{CERN-EP-2021-098}
 
\AtlasJournalRef{JHEP 01 (2022) 063}
 
\AtlasDOI{10.1007/JHEP01(2022)063}

\hypersetup{pdftitle={ANA-HDBS-2018-07-PAPER},pdfauthor={The ATLAS Collaboration}}
 
\begin{document}
 
\maketitle
 
\section{Introduction}
 
Since its introduction, the Standard Model (SM) has successfully
predicted several new particles, culminating in the discovery of the
Higgs boson (\Hboson)~\cite{HIGG-2012-27,CMS-HIG-12-028}.
The \Hboson boson, discovered by ATLAS and CMS at the CERN Large Hadron Collider (LHC)~\cite{Evans:2008zzb}, appears to have
properties consistent with those predicted by the SM within current
experimental
uncertainties~\cite{HIGG-2018-57,CMS-HIG-17-031,HIGG-2015-07}.
The
95\% confidence level (CL) upper limit on the branching ratio
for \Hboson boson decays to beyond-the-SM (BSM)
particles, from a
combined ATLAS and CMS measurement of the Higgs boson couplings, is
34\%~\cite{HIGG-2015-07},
although this limit comes with some assumptions. A more recent
ATLAS measurement,
based on 80~\ifb\
of 13~\TeV\ data~\cite{HIGG-2018-57},
set a 95\% CL upper limit of 21\% on the branching
ratio for \Hboson boson decays via undetected modes.
Given the magnitude of these limits and the assumptions that go into
deriving them, direct searches for exotic
decays of the Higgs boson remain a high priority
\cite{Curtin:2013fra}.
Among
the many final states discussed in Ref. \cite{Curtin:2013fra}, this
analysis most closely considers a scenario arising in
the next-to-minimal supersymmetric SM
(NMSSM)~\cite{Ellwanger:2009dp}, a generalization of the minimal
supersymmetric SM
(MSSM)~\cite{Fayet:1976et,Fayet:1977yc,Farrar:1978xj,Fayet:1979sa,Dimopoulos:1981zb}.

The MSSM is the simplest extension to the SM that incorporates supersymmetry (SUSY).
It predicts four additional Higgs bosons, generally assumed to be
heavier than the \Hboson boson: two neutral states, the $\Hzero$ and
$\Azero$ bosons, as well as two charged states, the \Hpm
bosons.
The measured mass of the
Higgs boson~\cite{HIGG-2016-33, CMS-HIG-19-004} close to
\SI{125}{\giga\electronvolt} results in the reintroduction of a
`little hierarchy' problem \cite{little_hierarchy} in the
MSSM. This hierarchy is alleviated in the
NMSSM by
allowing for additional contributions to the mass of the Higgs boson
from new scalar particles.  The NMSSM contains an
additional pseudoscalar Higgs boson ($a$), generally assumed to be
less massive than the \Hboson boson since its mass is protected by a
Peccei--Quinn (PQ) symmetry~\cite{Ellwanger:2009dp}.
 
Previous searches for exotic Higgs boson decays involving the
production of the $a$ boson have focused on the $R$-symmetry limit
\cite{Ellwanger:2009dp} of the
NMSSM, where the dominant decay channel is
\Hboson$\rightarrow{aa}$~\cite{HIGG-2014-02,HIGG-2017-09,HIGG-2017-05,HIGG-2016-03,HDBS-2018-47,CMS-HIG-14-019,CMS-HIG-17-029,CMS-HIG-17-024,CMS-HIG-18-011,CMS-HIG-18-006,CMS-HIG-18-024}.  As proposed
in Refs.~\cite{Huang:2013ima,Huang:2014cla}, the present analysis
considers instead the region of parameter space
near the PQ symmetry limit of the
NMSSM. Near this limit, the decay $\Hboson \rightarrow
\ninotwo\ninoone \rightarrow a \ninoone\ninoone$ dominates over
$\Hboson \rightarrow aa$, where \ninotwo and \ninoone are the two lightest
neutralinos, which are admixtures of the supersymmetric partners of
the Higgs and gauge bosons of the SM.  If the decay $a\rightarrow b\overline{b}$
is kinematically allowed, it is typically highly favoured. The \ninoone is assumed to be stable, as obtained in R-parity conserving SUSY models~\cite{Farrar:1978xj}.
 
The analysis presented in this paper uses \intlumi~\ifb\ of $pp$ collisions
at $\rts = 13$~\TeV\ collected with the ATLAS detector. It targets the \Hboson
$\rightarrow\ninotwo\ninoone\rightarrow{a}\ninoone\ninoone$
cascade decay, where the $a\rightarrow b\bar{b}$ decay dominates, and the
\Hboson boson is produced in association with a \Zboson boson.  The
\Zboson boson is required in the analysis selection
to decay into a pair of electrons or
muons (hereafter referred to as leptons), which provide a signature to
trigger upon and which reduce the multijet background.\footnote{The
contribution from $Z \rightarrow \tau^+\tau^-$ decays with the
subsequent decays of the $\tau$-leptons into light leptons is
included in the signal definition and simulation but is
suppressed by a factor of at least 2000 due to the event selection requirements.}
The resulting
final state consists of  a pair of oppositely charged
leptons, two jets, each containing a $b$-hadron (\bjets),
and missing transverse momentum (\met) from the two \ninoone neutralinos.  The
search is performed for a range of $m_a$ values, and for a few sets of fixed
values of $m_{\ninoone}$ and $m_{\ninotwo}$.
 
The primary Standard
Model backgrounds in this search are $Z$ bosons produced with
heavy-flavour (bottom and charm) jets, hereafter labelled \ZHF,
and \ttbar events; their contributions
are estimated from the data in control regions enhanced in
these backgrounds. The dijet invariant mass
is used as the final discriminant in a binned likelihood fit.
 
The search, which compares an expected background shape
in the signal region with the measured shape, is also sensitive in principle to other distortions of
the dijet invariant mass spectrum arising from BSM physics effects.

\section{The ATLAS detector}
 
\newcommand{\AtlasCoordFootnote}{
ATLAS uses a right-handed coordinate system with its origin at the nominal interaction point (IP)
in the centre of the detector and the \(z\)-axis along the beam pipe.
The \(x\)-axis points from the IP to the centre of the LHC ring,
and the \(y\)-axis points upwards.
Cylindrical coordinates \((r,\phi)\) are used in the transverse plane,
\(\phi\) being the azimuthal angle around the \(z\)-axis.
The pseudorapidity is defined in terms of the polar angle \(\theta\)
as \(\eta = -\ln \tan(\theta/2)\).
Angular distance is measured in units of \(\Delta R \equiv
\sqrt{(\Delta\eta)^{2} + (\Delta\phi)^{2}}\). The transverse momentum and transverse energy, \pT and \ET, are defined as $p\sin\theta$ and $E\sin\theta$, respectively.}
 
The ATLAS experiment~\cite{PERF-2007-01} at the LHC is a multipurpose particle detector
with a forward--backward symmetric cylindrical geometry and a near \(4\pi\) coverage in
solid angle.\footnote{\AtlasCoordFootnote}
It consists of an inner tracking detector (ID) surrounded by a thin superconducting solenoid
providing a \SI{2}{\tesla} axial magnetic field, electromagnetic and hadron calorimeters, and a muon spectrometer.
The inner tracking detector covers the pseudorapidity range \(|\eta| < 2.5\).
It consists of silicon pixel, silicon microstrip, and transition
radiation tracking detectors.  The insertable B-layer, installed
before Run~2~\cite{ATLAS-TDR-2010-19, PIX-2018-001}, typically provides
the innermost hit on a track.
Lead/liquid-argon (LAr) sampling calorimeters provide electromagnetic (EM) energy measurements
with high granularity.
An iron/scintillator hadron calorimeter covers the central
pseudorapidity range (\(|\eta| < 1.7\)).
The endcap and forward regions are instrumented with LAr calorimeters
for EM and hadronic energy measurements up to \(|\eta| = 4.9\).
The muon spectrometer surrounds the calorimeters and is based on
three large air-core toroidal superconducting magnets with eight coils each.
The field integral of the toroids ranges between \num{2.0} and \SI{6.0}{\tesla\metre}
across most of the detector.
The muon spectrometer includes a system of precision tracking chambers and fast detectors for triggering.
A two-level trigger system \cite{TRIG-2016-01} is used to select events.
The first-level trigger is implemented in hardware and uses a subset of the detector information
to keep the accepted event rate below \SI{100}{\kHz}.
This is followed by a software-based trigger that
reduces the accepted event rate to \SI{1}{\kHz} on average
depending on the data-taking conditions.
An extensive software suite~\cite{ATL-SOFT-PUB-2021-001} is used for real and simulated data reconstruction and analysis, for operation and in the trigger and data acquisition systems of the experiment.
 
\section{Data and simulated event samples}
\label{sec:samples}
The data used in this analysis, corresponding to the full Run 2
dataset for $pp$ collisions,  were collected at a centre-of-mass
energy of 13~\TeV\
during the 2015--2018 running periods using unprescaled single-lepton
triggers with a threshold of 26~\GeV\ (transverse energy, \ET,
for electrons and transverse momentum,
\pT, for muons) \cite{TRIG-2018-05, TRIG-2018-01}.
Events are selected for analysis only if they are of good quality
\cite{DAPR-2018-01} and
if all the relevant
detector components are known to have been in good operating
condition, which corresponds to a total integrated luminosity of
$139.0\pm 2.4$~\ifb~\cite{ATLAS-CONF-2019-021}.
The recorded events contain an average of 34 inelastic $pp$ collisions
per bunch-crossing.
 
\begin{table}[htbp]
\begin{center}{\fontsize{7}{7.9}\selectfont
\caption{Monte Carlo simulated samples used in this analysis. The top section
shows the nominal samples used for SM backgrounds and the potential
signal process.  The bottom section shows alternative SM samples for
evaluating the impact of theoretical systematic uncertainties.}
\begin{tabular}{l c c c c c}
\toprule
Process & Generator & Parton shower & PDF & Tune & Normalization \\
\midrule
\textbf{Nominal samples} & & & & & \\
\hsp \ttbar &
\POWHEGBOX v2~\cite{Frixione:2007nw,Nason:2004rx,Frixione:2007vw,Alioli:2010xd,ATL-PHYS-PUB-2016-020} &
\PYTHIA 8.230~\cite{Sjostrand:2014zea} &
\nnpdfnlo~\cite{Ball:2014uwa} &
A14~\cite{ATL-PHYS-PUB-2014-021}, \nnpdftwo~\cite{Ball:2012cx} &
NNLO+NNLL~\cite{Beneke:2011mq,Cacciari:2011hy,Baernreuther:2012ws,Czakon:2012zr,Czakon:2012pz,Czakon:2013goa,Czakon:2011xx} \\
\hsp \Zboson + jets &
\SHERPA 2.2.1~\cite{Bothmann:2019yzt} &
\SHERPA \cite{Hoeche:2011fd,Hoeche:2012yf,Catani:2001cc,Hoeche:2009rj} &
\nnpdfnnlo~\cite{Ball:2014uwa} &
\SHERPA &
NNLO~\cite{Anastasiou:2003ds} \\
\hsp Single-top ($Wt$) &
\POWHEGBOX v2 ~\cite{Re:2010bp}&
\PYTHIA 8.230 &
\nnpdfnlo &
A14, \nnpdftwo&
NLO+NNLL~\cite{Kidonakis:2010ux} \\
\hsp Diboson &
\SHERPA 2.2.1--2.2.2 &
\SHERPA &
\nnpdfnnlo &
\SHERPA &
NLO \\
\midrule
\hsp \multirow{2}*{NMSSM signal} &
\multirow{2}*{\POWHEGBOX v2} &
\multirow{2}*{\PYTHIA 8.210} &
\multirow{2}*{\cteq~\cite{Pumplin:2002vw}} &
\multirow{2}*{\aznlo~\cite{STDM-2012-23}} & NNLO(QCD) + \\
& & & & & NLO(EWK)~\cite{deFlorian:2016spz}\\
\midrule
\midrule
\textbf{Alternative samples} & & & & & \\
\hsp \ttbar &
\POWHEGBOX v2 &
\herwigseven~\cite{Bellm:2015jjp,Bahr:2008pv} &
\nnpdfnlo &
A14, \nnpdftwo &
NNLO+NNLL  \\
\hsp \multirow{2}*{\ttbar} &
\mgamc &
\multirow{2}*{\PYTHIA 8.230} &
\multirow{2}*{\nnpdfnlo} &
\multirow{2}*{A14, \nnpdftwo} &
\multirow{2}*{NNLO+NNLL} \\
& 2.6.0~\cite{Alwall:2014hca} & & & & \\
\hsp \multirow{2}*{\Zboson + jets} &
\mgamc &
\multirow{2}*{\PYTHIA 8.186~\cite{Sjostrand:2007gs}} &
\multirow{2}*{\nnpdfnnlo} &
\multirow{2}*{A14, \nnpdftwo} &
\multirow{2}*{NNLO}\\
& 2.2.2 & & & & \\
\bottomrule
\label{tab:samples}
\end{tabular}}
\end{center}
\end{table}
 
Although the dominant SM backgrounds are modelled using a data-driven technique, Monte Carlo (MC) simulated events provide input to these techniques and are used to model the subdominant backgrounds and the
$ZH$, \Hboson $\rightarrow \ninotwo\ninoone$ signal process.
A summary of all the MC event generator programs used in the analysis
is provided in Table~\ref{tab:samples}. Samples produced with
alternative generators are used to estimate systematic uncertainties
in the event modelling, as described in Section~\ref{sec:systematics}.
 
In the \Hboson$\rightarrow\ninotwo\ninoone$ simulated samples, a Higgs
boson is produced in
association with a \Zboson boson, using \POWHEGBOX, while \PYTHIA 8.210 is used to
force the decay chain: $\Hboson\rightarrow\ninotwo\ninoone$
and $\ninotwo\rightarrow{a}\ninoone$.  The \Zboson boson is forced
to decay into $e^+e^-$ or $\mu^+\mu^-$ or
$\tau^+\tau^-$.  Both the Higgs and $a$
bosons have narrow widths, with the Higgs boson width set to its SM value
and the $a$ width set to its mass (in \GeV) times $10^{-5}$.  The $a$
boson is then required to decay into a pair of \bquarks. The $a$ width
is narrow enough that the
experimental resolution dominates the reconstructed dijet invariant mass width for all masses considered here.

All simulated processes are normalized using the most accurate
theoretical cross-section predictions currently available and were
generated at least
to next-to-leading-order QCD accuracy.  All
samples of simulated background events were passed through the ATLAS detector
simulation~\cite{SOFT-2010-01} based on
$\GEANT$~\cite{Agostinelli:2002hh},
while signal samples were passed through a fast simulation~\cite{ATL-PHYS-PUB-2010-013}
based on a parameterization of showers in the ATLAS calorimeters
and employing $\GEANT$ elsewhere.
The effects
of multiple interactions in the same and nearby bunch crossings
(pile-up) were modelled by overlaying the hard-scatter events with minimum-bias events
simulated using the soft QCD processes of \textsc{Pythia~8.186}~\cite{Sjostrand:2007gs}
with the A3~\cite{ATL-PHYS-PUB-2016-017}
set of tuned parameters (tune) and
\textsc{NNPDF2.3LO}~\cite{Ball:2012cx} parton distribution
functions (PDF). The minimum-bias samples were reweighted such that
the pile-up distribution matches that in the data.
For all samples of simulated events, except for
those generated using \textsc{Sherpa}~\cite{Bothmann:2019yzt}, the
\textsc{EvtGen~1.6.0} program~\cite{EvtGen} was used to
describe the decays of bottom and charm hadrons.
 
\section{Object and event selection}
 
\subsection{Object reconstruction}
 
Tracks measured in the ID \cite{PERF-2015-08} are used to
reconstruct interaction
vertices~\cite{ATL-PHYS-PUB-2015-026}, of which the one with the
highest sum of squared
transverse momenta of associated tracks is selected as the primary
vertex of the hard interaction.
 
Electrons are reconstructed from clusters of energy
deposits~\cite{PERF-2014-07}  in the electromagnetic
calorimeter and matched to a track in the
ID~\cite{EGAM-2018-01}.
One electron must satisfy the {\em Tight}
identification criteria with $\ET > 30$~\GeV\ to be on the trigger
efficiency plateau, and be matched to the trigger electron, while the second is required to satisfy the {\em Medium} identification
criteria with $\ET > 20$~\GeV~\cite{EGAM-2018-01}. In addition, these
electrons must have $|\eta |<2.47$, be outside the transition region
between the barrel and endcap calorimeters ($1.37 < |\eta| < 1.52$),
and have small impact
parameters:\footnote{Transverse (\dzero) and longitudinal ($z_{0}$)
impact parameters are
defined relative
to the primary vertex position, where the beam line is used to
approximate the primary  vertex position in the transverse plane. The uncertainty in \dzero is denoted by $\sigma_{d_{0}}$.}
$|\dzero /\sigma_{d_{0}}| < 5$
and $|\zzsth| < 0.5$~mm.
Finally, these electrons are required to pass the {\em Gradient} isolation
requirements~\cite{EGAM-2018-01}.
 
Muons are reconstructed as described in Refs.~\cite{PERF-2015-10,MUON-2018-03} and
required to have $|\eta|<2.7$. The
leading \pt muon is required to have $\pt > 30$~\GeV\ so as to be on the
trigger efficiency plateau, and must be matched to the trigger muon, while the second muon is required to have
$\pt > 20$~\GeV.
These muons must satisfy the {\em Medium} identification criteria,
pass the {\em Gradient} isolation
requirements~\cite{PERF-2015-10}, and have $|\dzero /\sigma_{d_{0}}| < 3$
and $|\zzsth| < 0.5$~mm.
 
In order to veto events with additional leptons, looser
selection criteria are employed.  Electrons are required to pass
the {\em LooseAndBLayerLLH} requirements with $\ET > 10$~\GeV\ and
$|\eta|<2.47$. Muons are required to satisfy the {\em Medium} criteria
with $\pt > 4$~\GeV\ and $|\eta| < 2.7$.
 
Jets are reconstructed from energy deposits in clusters of calorimeter
cells \cite{PERF-2014-07}
using the \antikt\ algorithm~\cite{Cacciari:2008gp,Fastjet} with radius
parameter $R = 0.4$.  Jet cleaning criteria are used to identify jets
arising from non-collision backgrounds or noise in the
calorimeters~\cite{ATLAS-CONF-2015-029} and events containing such
jets are removed. Jets are calibrated using the standard energy scale corrections~\cite{PERF-2016-04}.
Jets are required to have $\pt>20$~\GeV\ and $|\eta|<2.4$.
A jet vertex tagger~\cite{PERF-2014-03} at the {\em Medium} working
point is used to remove jets with $20 < \pt < 60$~\GeV\ and
$|\eta|<2.4$
which are identified as not being
associated with the primary vertex of the hard interaction.
Jets containing a $b$-hadron are identified as \bjets ($b$-tagged)
using the MV2 multivariate discriminant~\cite{FTAG-2018-01}, with
the selection tuned to produce an average efficiency
of 77\% for \bjets, with corresponding
light-flavour ($u$-, $d$-, $s$-quark and gluon) and $c$-jet
misidentification efficiencies of 0.9\% and 25\% respectively, as
measured in
simulated \ttbar events.

Jets and leptons are reconstructed independently. To prevent
double counting of these reconstructed objects,
an overlap removal procedure for leptons and
jets is applied \cite{SUSY-2016-15}.  The looser selection criteria for leptons,
as described above, are employed for the overlap removal.
 
The missing transverse momentum, with magnitude
\Etmiss, is calculated as the negative vector sum of the transverse
momenta of all calibrated selected objects, such as electrons and
jets, and is corrected to take into account the transverse momentum of
muons. Tracks with $\pt > 500$~\MeV, compatible with the primary vertex
but not matched to any reconstructed object, are included in the
\Etmiss\ reconstruction to take into
account the soft-radiation component that does not get clustered into
any hard object \cite{PERF-2016-07}.
 
To account for small efficiency differences between simulation and
data, simulated events are corrected  with scale factors covering
lepton reconstruction, identification, isolation and trigger
efficiencies, as well as jet pile-up rejection and flavour tagging efficiencies.

\subsection{Event selection}
 
To select events consistent with the decay of the Higgs boson into a
$b\bar{b}$ pair plus \Etmiss, this analysis focuses on $ZH$ production
in which the leptonic decay of the \Zboson boson into electrons or muons
provides the trigger signature for the event.

Events are required to have two leptons of the same flavour and
opposite charge; events are rejected if any additional leptons are
found.  Events containing muons that are poorly reconstructed, with
$\sigma(q/p)/|q/p| > 0.2$ where $q/p$ is the charge-to-momentum ratio, or muons from cosmic-ray background,
with $|\dzero| > 0.2$~mm or $|z_{0}| > 1$~mm, are also rejected.
 
Events in the signal region (SR) are required to satisfy the following:
\begin{itemize}
\item Dilepton invariant mass in the range $81 < \mll < 101$~\GeV\
\item Dilepton $\pt$ with $\ptll > 40$~\GeV\
\item At least two jets with $\pt >20$~\GeV
\item $\Etmiss > 100$~\GeV\
\item Dijet invariant mass in the range $20 < \mjj < 120$~\GeV, based on the two jets
with the highest \pt in the event, at least one of which must be $b$-tagged
\item A \pt fraction (\ptfrac), defined \cite{Huang:2013ima} as the scalar sum of the \pt of the dijet system and \Etmiss\ divided by the dilepton \pt, in the range:
\begin{equation*}
0.8 < \frac{\pt^{jj}+\Etmiss }{\ptll} < 1.2
\end{equation*}
\end{itemize}
 
The requirement on \ptfrac\ is especially useful in reducing the \ttbar
background. Requiring only one $b$-tagged jet is a trade-off between signal acceptance and background rejection.
Although the dijet resonance search is mainly sensitive to decays of
particles with masses in the range 20--65~\GeV, the window for the dijet
invariant mass extends to higher values to take into account the tail
in the dijet invariant mass distribution that results from choosing a jet that does not come from $a\rightarrow b\overline{b}$, and also to better constrain the background shape.
 
In addition to the SR, two control regions are defined.
A control region for \ZHF (\CRZ) is defined with the same
requirements as the SR except with $60 < \Etmiss < 100$~\GeV. A \ttbar
control region (\CRTop) is defined with the same criteria
as the SR but with the \mll
requirement inverted (and $\mll > 50$~\GeV).  To validate the modelling of
\Etmiss\ for the \ZHF processes, a validation region VRMET is defined
with the same criteria as the SR except that the \Etmiss\
requirement is loosened to $\Etmiss > 50$~\GeV\ and the dijet invariant
mass requirement is $\mjj > 150$~\GeV.
 
The selection criteria for the signal, control and validation regions
are summarized in Table~\ref{t:event_selection}. The acceptance times efficiency (for $\Hboson \rightarrow \ninotwo\ninoone \rightarrow a
\ninoone\ninoone \rightarrow b\bar{b} \ninoone\ninoone$)
of the selection varies across the three dimensions of $m_a$, $m_{\ninoone}$, and $m_{\ninotwo}$, but is primarily a function of $m_a$ and varies from approximately 0.4\% to 1.1\%, depending on the configuration of the three masses.
 
\begin{table}[htbp]
\begin{center}{\fontsize{11}{13.2}\selectfont
\caption{Summary of the event selection criteria for the signal region
(SR), $Z$+jets control region (CRZ), \ttbar control region (CRTop), and
a validation region for \met modelling in \ZHF events
(VRMET). The first five selection criteria are common to all
regions.}
\begin{tabular}{l c c c c}
\toprule
& SR & CRZ & CRTop & VRMET \\
\midrule
Number of leptons & \multicolumn{4}{c}{~~~~2} \\
Number of jets & \multicolumn{4}{c}{$\ge 2$} \\
Number of $b$-tagged jets & \multicolumn{4}{c}{$\ge 1$} \\
Dilepton \pt[\GeV] & \multicolumn{4}{c}{$> 40$~~~} \\
\pt~fraction & \multicolumn{4}{c}{[0.8, 1.2]} \\
\midrule
Dilepton mass [\GeV] & [81, 101] & [81, 101] & [50, 81] or $>101$ & [81, 101]  \\
\met~[\GeV] & ~~$> 100$ & [60, 100] & ~~$> 100$ & ~~~~$> 50$ \\
Dijet mass [\GeV] & [20, 120] & [20, 120] & [20, 120] & ~~$> 150$ \\
\bottomrule
\label{t:event_selection}
\end{tabular}}
\end{center}
\end{table}

\section{Background and statistical model}

The background in the SR is primarily composed of \ZHF and
\ttbar, with a small contribution from $Z$ plus light-flavour jets ($Z$+light),
single-top and diboson events. The background from multijet production was studied using same-sign lepton pairs and found to be negligible.
The cross sections for \ZHF and \ttbar production
in the simulation are scaled by normalization factors,
$\mu_{\ZHF}$ and $\mu_{\ttbar}$, respectively, determined from a simultaneous
fit 
to the number of events in data and simulation
in \CRZ and \CRTop. Subdominant backgrounds in
these regions are normalized to the theoretical cross sections
specified in Section \ref{sec:samples}, while the \ZHF and \ttbar components are allowed to float. The calculated scale
factors are $\mu_{\ZHF} = 0.955\pm 0.032$ and $\mu_{\ttbar} = 0.798 \pm
0.033$ where the uncertainties are statistical.
Figure~\ref{fig:dataMC} shows
comparisons between data and simulation in the control and validation
regions for a few characteristic
observables after applying the \ZHF and
\ttbar scale factors.
The
contribution of signal events is less than 1\% in \CRTop and at most 3\% in \CRZ for all signal points.
 
\begin{figure}[htbp]
\begin{center}
\subfloat[]{
\includegraphics[width=0.45\textwidth]{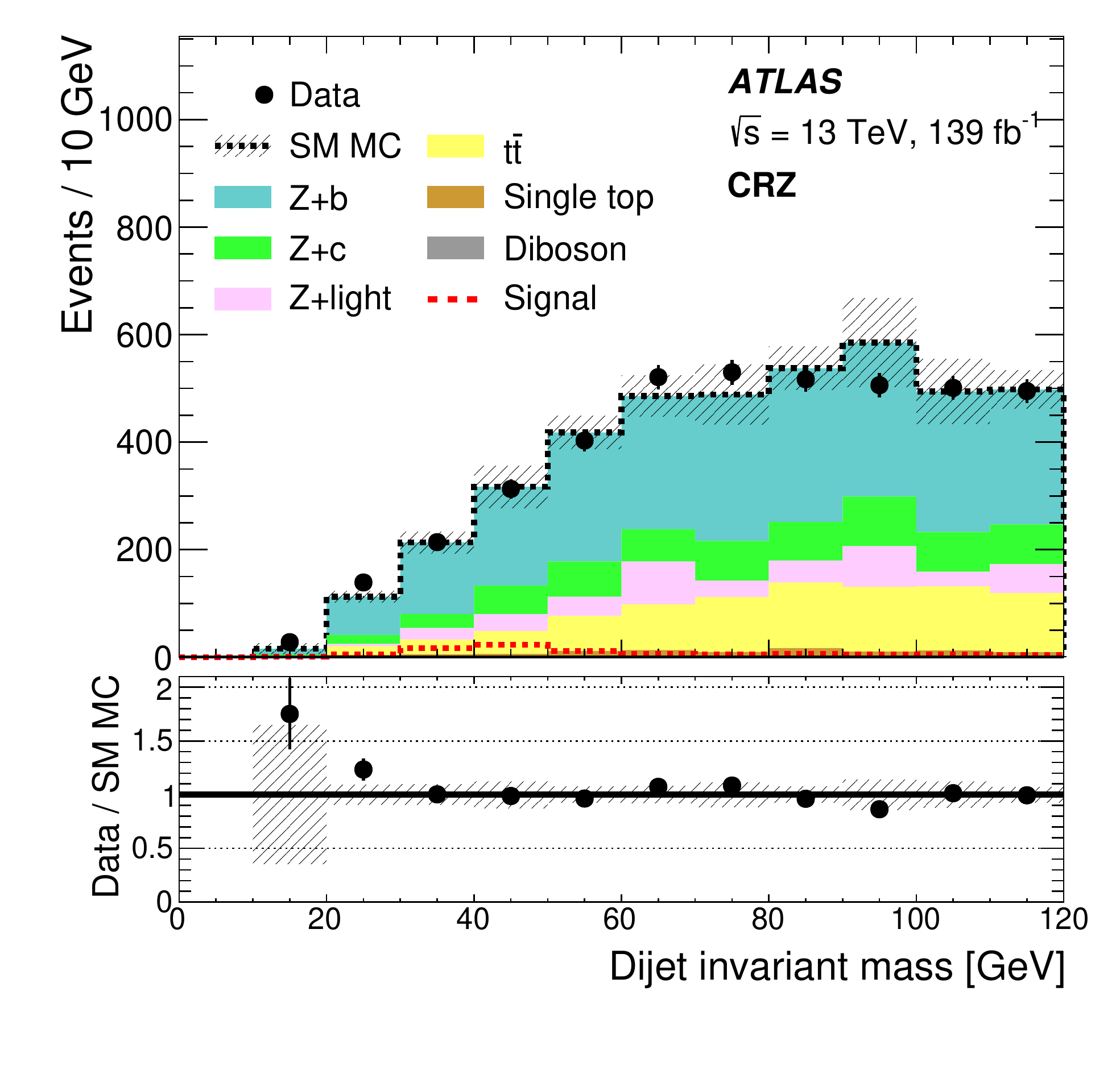}
}
\subfloat[]{
\includegraphics[width=0.45\textwidth]{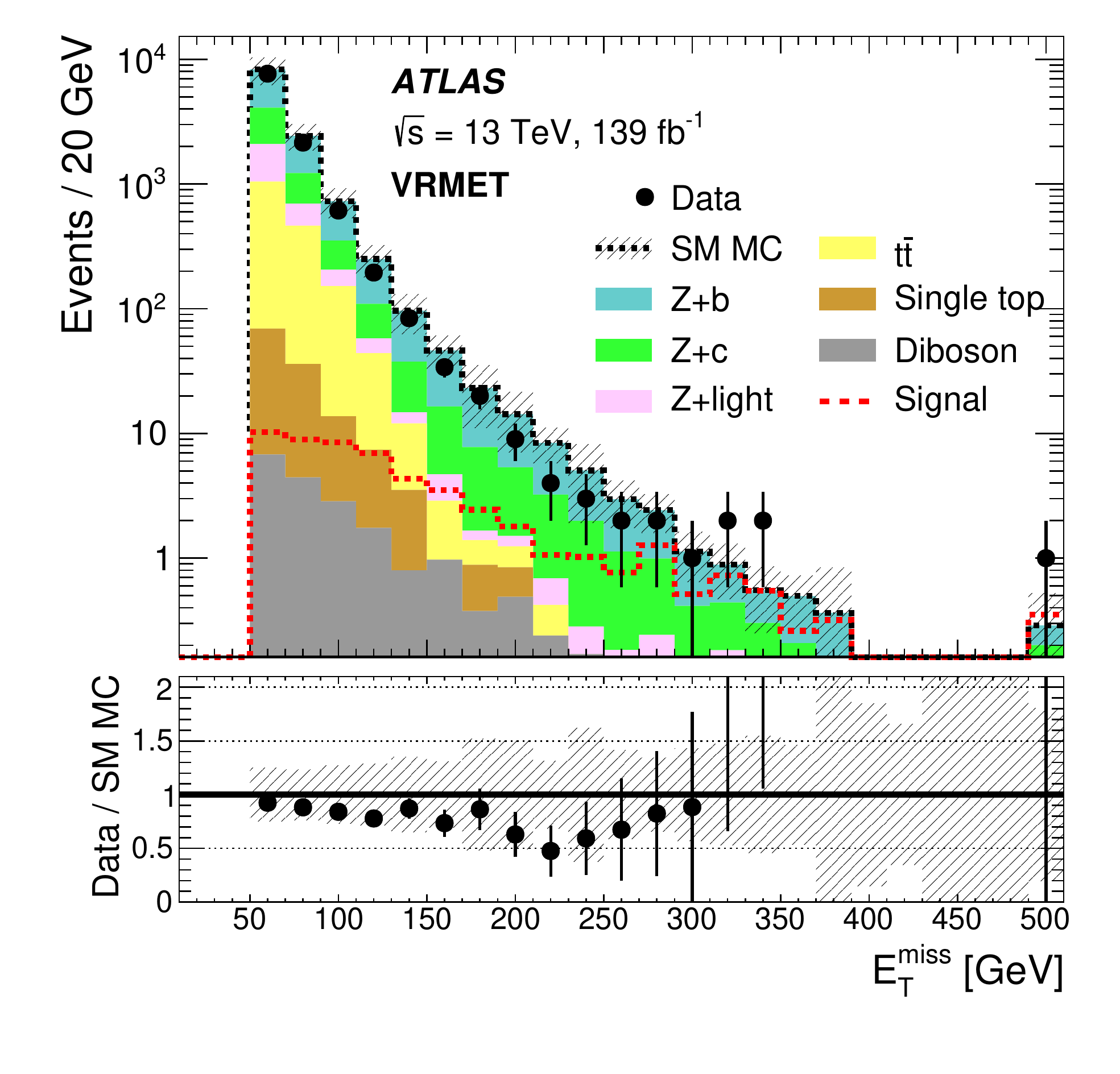}
}
\\
\subfloat[]{
\includegraphics[width=0.45\textwidth]{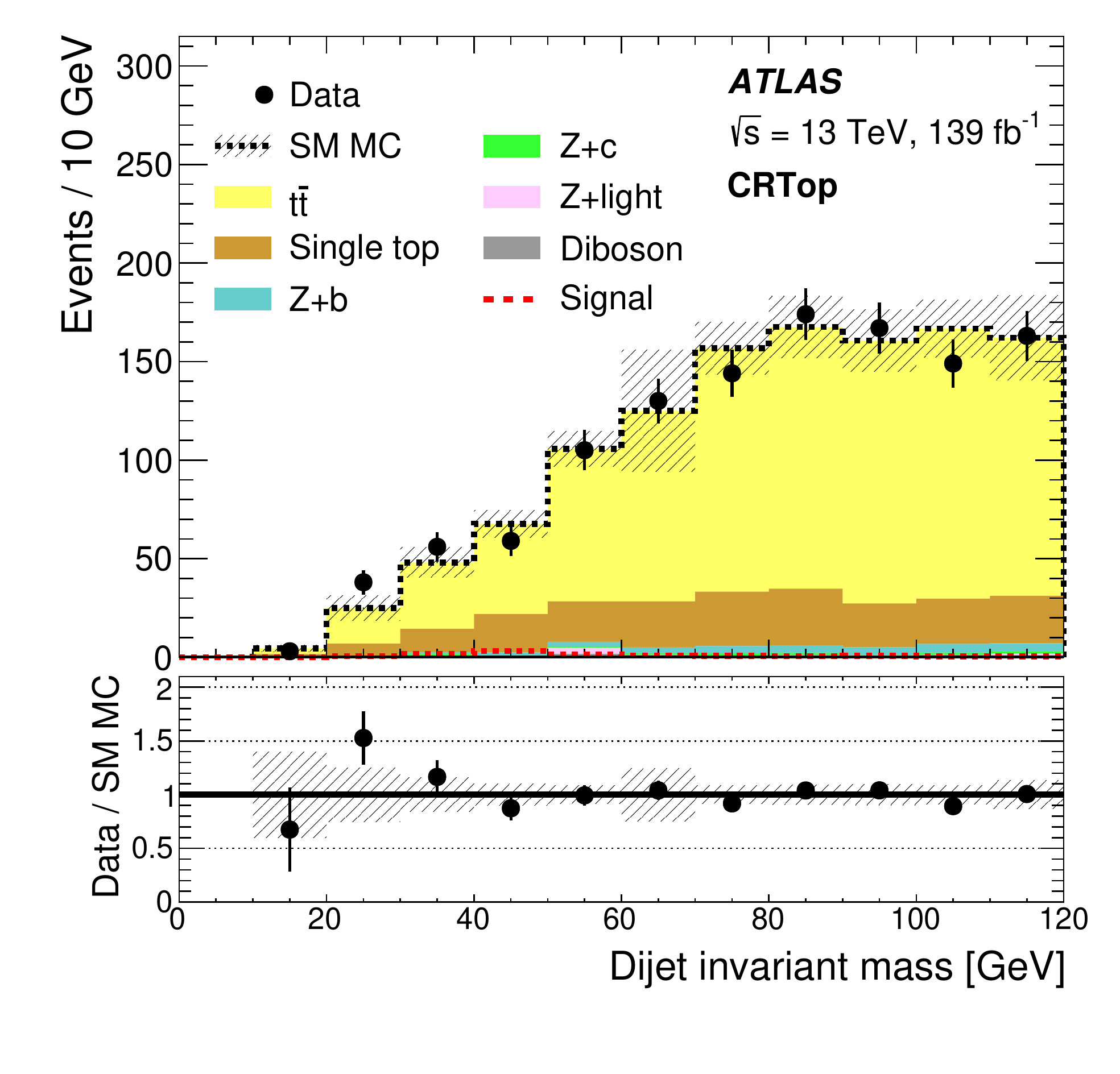}
}
\subfloat[]{
\includegraphics[width=0.45\textwidth]{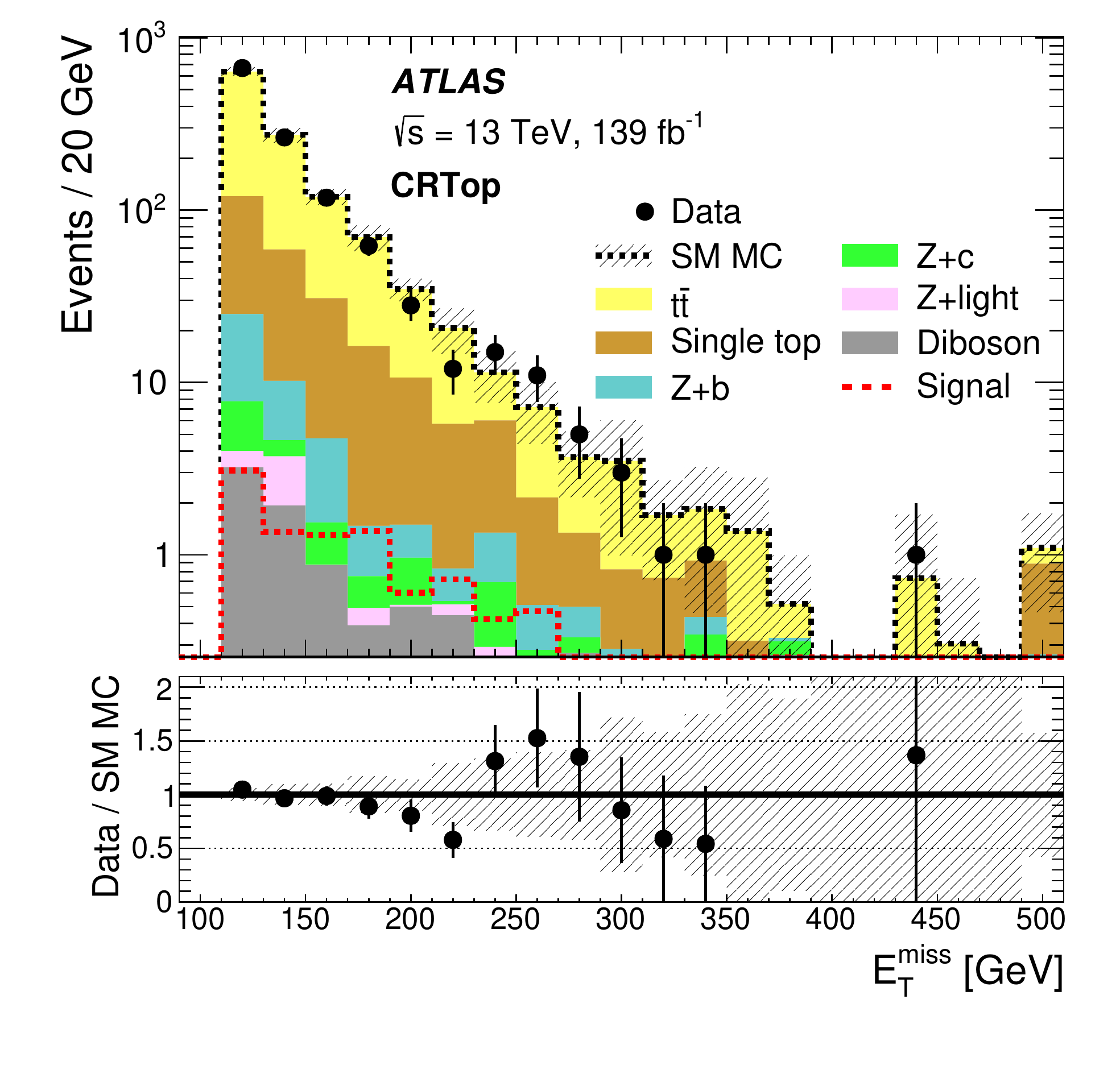}
}
\caption{Comparison of data and simulation for (a) the dijet invariant
mass in \CRZ, (b) \Etmiss\ in \VRMET, (c) the dijet invariant
mass in \CRTop and (d) \Etmiss\ in \CRTop. The \ZHF and \ttbar scale factors, described in the text, have been applied to the simulated samples.
The lower
panels show the ratio of data to SM MC simulation.  The total
statistical and systematic uncertainties are denoted by the
hatched band. Large fluctuations in single bins of the systematic uncertainty band can be caused by high-weight events in the systematic variation samples. The overlaid
distribution labelled `Signal' is for the model with ($m_a$,
$m_{\ninoone}$, $m_{\ninotwo}$) = (45~\GeV, 10~\GeV, 80~\GeV),
setting all branching ratios to 100\% in the decay chain
$\Hboson \rightarrow \ninotwo\ninoone \rightarrow a
\ninoone\ninoone \rightarrow b\bar{b} \ninoone\ninoone$.}
\label{fig:dataMC}
\end{center}
\end{figure}
 
As described below, the background in the SR is modelled using a combination of the
\mjj distribution shapes from data in \CRZ and \CRTop to model \ZHF and \ttbar,
respectively, with only a small dependence on the MC
simulation. The relatively small
contributions of non-\ZHF events to \CRZ (approximately 30\%) and of non-\ttbar events to \CRTop (approximately 20\%)
are derived from MC simulation and subtracted, after applying the scale factors $\mu_{\ZHF}$ and $\mu_{\ttbar}$ to the simulated data.
The resulting \mjj distributions, assumed to correspond
to pure \ZHF and \ttbar, are referred to as the `\CRZ \ZHF' and
`CRTop \ttbar' distributions.

The shape of the \mjj distribution for \ZHF in \CRZ is found to
differ slightly from that in the SR, according to the predictions
of the \SHERPA MC generator. Therefore, the bin-by-bin ratio
of the \mjj distributions in \CRZ and the SR, obtained from MC simulation
and denoted by $\UZHF$, is applied
to the \CRZ \ZHF \mjj distribution from the data when assembling the background
model. The correction is linear as a function
of \mjj and ranges from a factor of 0.6 at 20~\GeV\ to just under 1.2 at 120~\GeV.
A similar comparison via MC simulation between the \mjj
shapes in \CRTop and the SR shows no statistically significant shape
difference. Therefore, no shape correction is made for \CRTop.
 
The background model is constructed from the data as a weighted sum of the
\CRZ \ZHF and \CRTop \ttbar distributions to which is
added the
\mjj distribution of the subdominant backgrounds in the SR obtained
from simulation. Signal contamination in the CRs is neglected. The
weights for the \CRZ and \CRTop components, denoted \RZHF\ and
\Rttbar\ respectively, are the ratios of the integral event counts
from simulation
between \(\mjj = \SI{20}{\GeV}\) and \(\SI{120}{\GeV}\) in the \SR
to those in the respective control region.
The background model can be described as
\begin{equation}\label{eq:background_model}
\mubkg \NmodelSR~ \fmodelSR = \mubkg \left(\NsubdataCRTop~
\Rttbar  \fsubdataCRTop + \NsubdataCRZ \RZHF \fsubdataCRZcorr + \NsubdomSR~ \fsubdomSR \right), \\
\end{equation}
where \(N_{j}^{i}\) is the total event count of sample \(i\) in region \(j\), and \(\hat{\mathbf{f}}_{j}^{i}\) is the probability density function
as a function of \mjj of sample \(i\) in region \(j\), both
obtained from data in the respective control regions.  In particular, $\fsubdataCRZcorr$ is the normalized product of \(\mathbf{U}_{\text \ZHF}\) and $\fsubdataCRZ$, where \(\mathbf{U}_{\text \ZHF}\) is the \mjj bin-by-bin shape correction factor described above. The superscript `subdom.'\ on the last term in \Eqn{(\ref{eq:background_model})} indicates the subdominant backgrounds ($Z$+light, single-top, and diboson events) that are derived directly from the Monte Carlo simulation in the signal region, as indicated by the subscript `SR-MC'.
 
Because \(R_i\) and \(\mathbf{U}_{Z+HF}\) are
ratios of quantities from the MC samples, they are less
sensitive to detector and theory systematic uncertainties.  The factor
\mubkg is a background normalization parameter whose expected value is close to unity, given that
the left-hand side is normalized to \NmodelSR, the number of
background events expected from simulation in the SR, after having
applied the \ZHF and \ttbar scale factors, $\mu_{\ZHF}$ and $\mu_{\ttbar}$, respectively.
To smooth out statistical fluctuations in the background shape, the
histogram resulting from the superposition in
\Eqn{(\ref{eq:background_model})} is fit with a fourth-order polynomial
over the range \(\mjj \in (\SI{20}{\GeV}, \SI{120}{\GeV})\).
 
Hypothesis tests for the presence of an exotic Higgs boson decay signal are
performed with a
multi-bin fit of the $m_{jj}$ distribution in the SR, based on a
product of Poisson likelihood terms in a
profile likelihood-ratio \cite{Cowan:2010js} test statistic.  Systematic
uncertainties are included as additional terms in the likelihood,
assuming Gaussian auxiliary measurements for the nuisance
parameters. Upper limits are obtained at 95\% CL,
based on the CL$_\textrm{s}$ prescription \cite{Read:2002hq}.
 
The model that is fit to the data is the sum of two \mjj templates:
\begin{itemize}
\item The SM background, whose shape comes from the parameterization
described earlier in this section.  The
normalization is controlled by the nuisance parameter \mubkg.
\item The signal, whose shape comes directly from simulation.
The signal normalization is controlled by the fit parameter-of-interest,
\musig, which is constrained to be positive.  The parameter is scaled
to correspond to the branching ratio for the exotic Higgs boson decay, \Hboson $\rightarrow\ninotwo\ninoone\rightarrow{a}\ninoone\ninoone$, without constraining the branching ratio to be less than unity, and
assuming the Standard Model cross section for $ZH$
associated production and a branching ratio BR$(a \rightarrow b\bar{b}) = 100$\%.
\end{itemize}
 
\section{Systematic uncertainties}
\label{sec:systematics}
Systematic uncertainties affect our ability to model the shape of the background dijet invariant mass spectrum from simulation, and therefore the background model used
in the final fit. Since the background model is built
primarily using the data \mjj distributions in \CRZ and \CRTop, the
impact of several sources of theoretical and experimental uncertainty
is reduced. Nevertheless,
the simulated \mjj distributions enter through the subdominant
backgrounds in both control regions and in the SR, through the
predicted yields of \ZHF and \ttbar in the three regions, and
through the \mjj shape correction for \ZHF events, all of which
enter into \Eqn{(\ref{eq:background_model})}. In addition, the signal
acceptance and \mjj line shape are subject to theoretical and
experimental uncertainties.
 
Systematic uncertainties are evaluated for detector effects, for
theoretical modelling and for variations in the parameterized
background model shape
arising from the statistical uncertainties in the polynomial fit
parameters. In
addition, a systematic uncertainty is evaluated to account for the
choice of background functional form by comparing the results obtained using the fourth-order
polynomial with those from a perfect fit to the data, i.e.\ the
background-model data points
themselves. The uncertainty associated with a systematic effect is
taken into account by producing a new background \mjj distribution that includes
the systematic variation. Scale factors for \ZHF and \ttbar and the
\mjj shape correction for \ZHF are recomputed for each systematic
variation and the background analysis is fully re-run to determine the size
of the effect on the shape of the background model.
 
The procedure to derive the background model introduces statistical
uncertainties when subtracting the subdominant backgrounds. These
uncertainties are accounted for in the statistical uncertainty of the
background model.
However, to avoid double counting of these statistical uncertainties
when evaluating the impact of systematic variations,
the subtraction of the subdominant backgrounds is omitted; both the nominal
distribution and the systematically varied distribution are evaluated
without subtracting the subdominant backgrounds.  Comparing the shape
of the background model with and without the subtraction of the
subdominant backgrounds, the difference, measured as the maximum absolute difference between the two background curves, was found to be at most 3\%.
 
Detector-related uncertainties are evaluated for the jet energy scale
and resolution \cite{PERF-2016-04}, the efficiency of the jet vertex
tagger~\cite{PERF-2014-03}, the $b$-tagging performance
\cite{FTAG-2018-01,ATLAS-CONF-2018-006,ATLAS-CONF-2018-001}, pile-up reweighting, and lepton
reconstruction efficiency, energy/momentum scale, and resolution
effects \cite{EGAM-2018-01,PERF-2015-10}.
The jet- and lepton-related uncertainties
are propagated to the calculation of \Etmiss; an additional
uncertainty for the soft-radiation component contributing to \Etmiss\
is also taken into account \cite{PERF-2016-07}. The
uncertainty in the combined 2015--2018 integrated luminosity is 1.7\%
\cite{ATLAS-CONF-2019-021}, obtained using the LUCID-2 detector
\cite{Avoni:2018iuv} for the primary luminosity measurements.
 
Uncertainties in the theoretical modelling of the dominant background
processes (\ZHF and \ttbar) are evaluated as follows.  For
\ZHF, the effects of varying the factorization and renormalization scales were
evaluated using seven variations in \SHERPA
\cite{Bothmann:2016nao} each of which changed
the renormalization and factorization scales by a factor of 0.5 or 2,\footnote{The pairwise variations
applied to the renormalization (\muR) and factorization (\muF)
scales are $[\muR, \muF] \times [0.5,0.5], [1,0.5], [0.5,1], [1,1],
[2,1], [1,2], [2,2]$.} leading to a maximum shape difference in the \mjj distribution of 4\%.  Results are also evaluated using an
alternative sample based on \mgamc \cite{Alwall:2014hca} and
their difference from those of the nominal \SHERPA samples, at most 5\%, is assigned as an additional systematic uncertainty.
 
For \ttbar production, uncertainties are estimated by comparing
different matrix-element calculations (\POWHEGBOX vs \aMCatNLO),
different choices of parton-showering model
(\pythia vs \herwigseven), and different amounts of initial-state (ISR) and final-state (FSR) radiation within \pythia, while leaving all other parameters for
each comparison unchanged. The uncertainty due to ISR is estimated by
comparing the nominal \ttbar\ sample with two additional
samples to simulate higher/lower parton
radiation~\cite{ATL-PHYS-PUB-2017-007}. The impact of
FSR is evaluated by varying the
renormalization scale for emissions from the parton shower up and down
by a factor of two. All of these variations lead to differences compared to the nominal choice of less than 2\%.
 
The systematic uncertainty in the $ZH$ cross section is taken from
Ref.~\cite{deFlorian:2016spz} and includes the effect of QCD scale,
PDF, and \alphas uncertainties.  The impact of theoretical
uncertainties on the Higgs boson production kinematics
were neglected, based on the studies described in Ref. \cite{HIGG-2018-04}.

\section{Results}
 
The \mjj distribution in the SR is shown in
Figure \ref{fig:money_plot}.  The observation is consistent with the
background model for the SM hypothesis. The long tail on the example
signal distribution comes from selecting a jet that does not
come from the $a\rightarrow b\bar{b}$ decay. For the signal models
considered below, which include $m_a$ from 20~\GeV\ to 65~\GeV, the smallest $p$-value \cite{Cowan:2010js}
is 0.39 for the model with
($m_a$, $m_{\ninoone}$, $m_{\ninotwo}$) = (50~\GeV, 10~\GeV, 110~\GeV).

\begin{figure}[h]
\begin{center}
\includegraphics[width=0.60\textwidth]{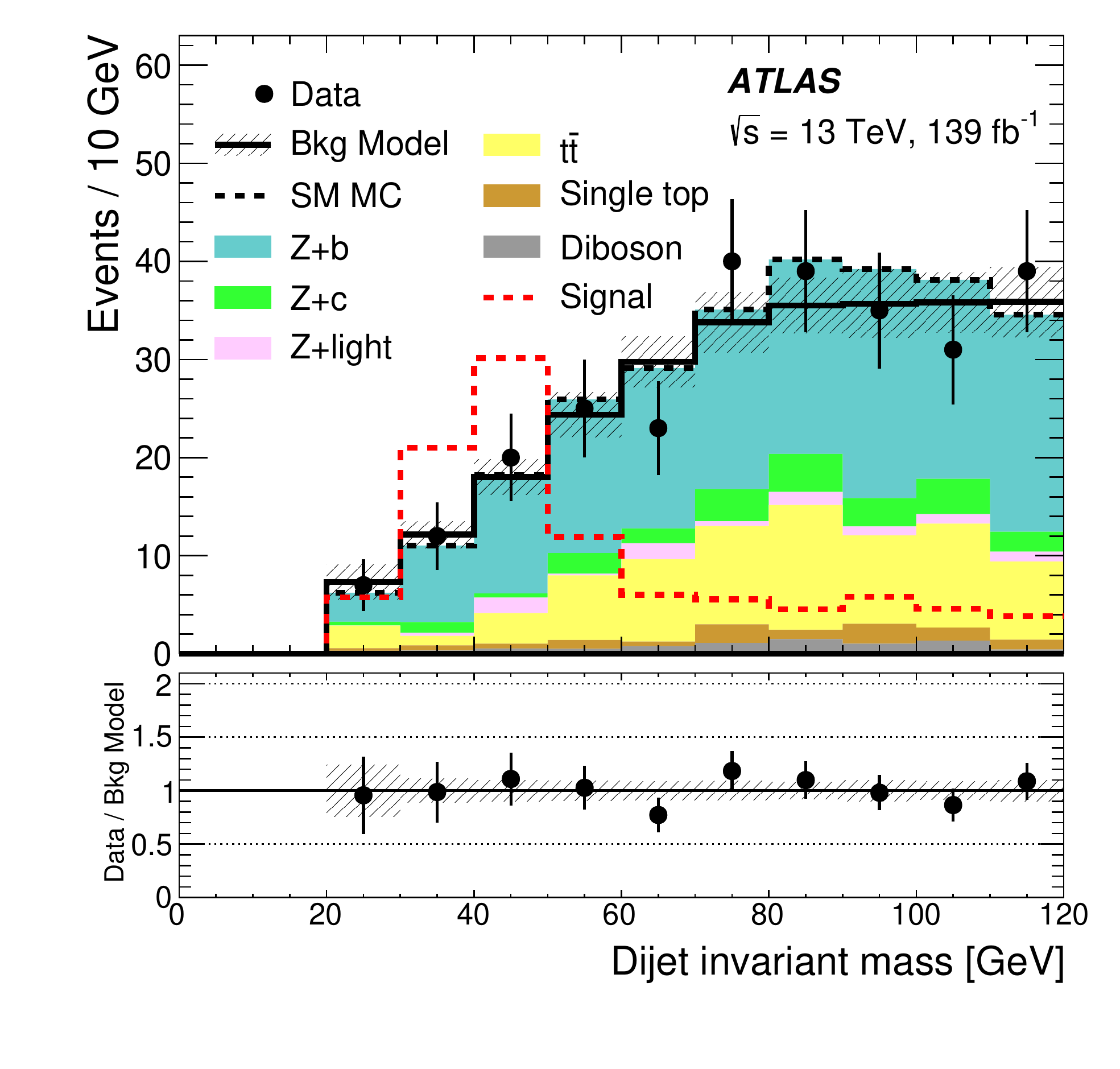}
\caption{Distribution of the dijet invariant mass
in the signal region, shown together with the
parameterized background model (labelled `Bkg Model'). For reference, the MC prediction for the SM background is also shown (labelled `SM MC'). The \ZHF and \ttbar scale factors, described in the text, have been applied to the simulated samples.  The
signal region is defined to have dijet invariant mass $> 20$~\GeV.
The total statistical and systematic uncertainties are denoted
by the hatched band. The
distribution labelled `Signal' is for the model with ($m_a$,
$m_{\ninoone}$, $m_{\ninotwo}$) = (45~\GeV, 10~\GeV, 80~\GeV),
setting all branching ratios to 100\% in the decay chain
$\Hboson \rightarrow \ninotwo\ninoone \rightarrow a
\ninoone\ninoone \rightarrow b\bar{b} \ninoone\ninoone$. The long tail on the signal distribution comes from selecting a jet that does not come from the $a\rightarrow b\bar{b}$ decay. The lower
panel shows the ratio of the observed data to the expectation
from the parameterized background model. }
\label{fig:money_plot}
\end{center}
\end{figure}
 
The dominant uncertainty limiting the final result is statistical
and comes from the limited number of events in the signal region.  The dominant
systematic uncertainty is again statistical in
nature and comes from the normalization of the background shape
function to the data observed in the SR; for signal models where this
analysis has sensitivity, this uncertainty has a 6\%--10\% effect on the
fitted \musig.
Subdominant systematic uncertainties in this
analysis include the theoretical uncertainty of the \ZHF \mjj shape
correction
(2\%--3\%), the jet energy resolution (1\%--3\%), the flavour dependence of the
jet energy scale (1\%--2\%), and
the statistical uncertainties of the background shape parameters (1\%--3\%),
which arise from the limited number of data events in \CRZ and \CRTop.
 
Upper limits at 95\% CL on the $pp \rightarrow \ZH$ cross section times branching ratio
for $Z \rightarrow \ell^{+}\ell^{-}$ (where $\ell = e, \mu~
\mathrm{or}~ \tau$) and $\Hboson \rightarrow
\ninotwo\ninoone \rightarrow a \ninoone\ninoone \rightarrow b\bar{b}
\ninoone\ninoone$ are shown in Figure
\ref{fig:brazil_plot} as a function of $m_{a}$ for several fixed values of
$m_{\ninoone}$ and $m_{\ninotwo}$.  Assuming the SM value for \ZH\
production, these results can be interpreted as upper
limits on the branching ratio for $\Hboson \rightarrow
\ninotwo\ninoone \rightarrow a \ninoone\ninoone \rightarrow b\bar{b}
\ninoone\ninoone$.  In the region
of highest sensitivity, a branching ratio upper limit of 31\% is obtained. For comparison between the results at different $\ninotwo$, $\ninoone$ mass points, Figure \ref{fig:UL_all} shows the six 95\% CL upper limits of Figure \ref{fig:brazil_plot} together, without the uncertainty bands.
 
\begin{figure}[htbp]
\begin{center}
\includegraphics[width=0.45\textwidth]{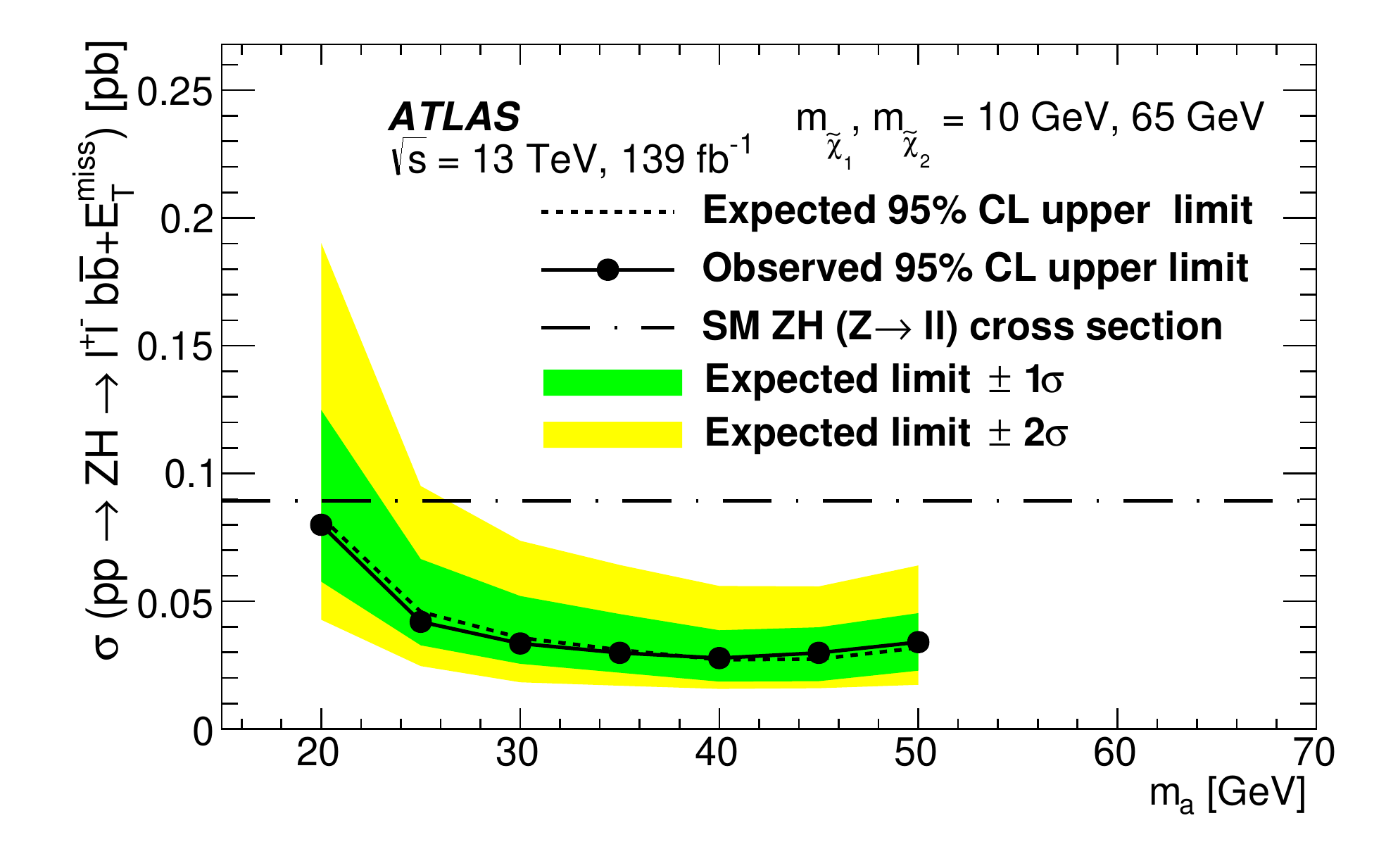}
\includegraphics[width=0.45\textwidth]{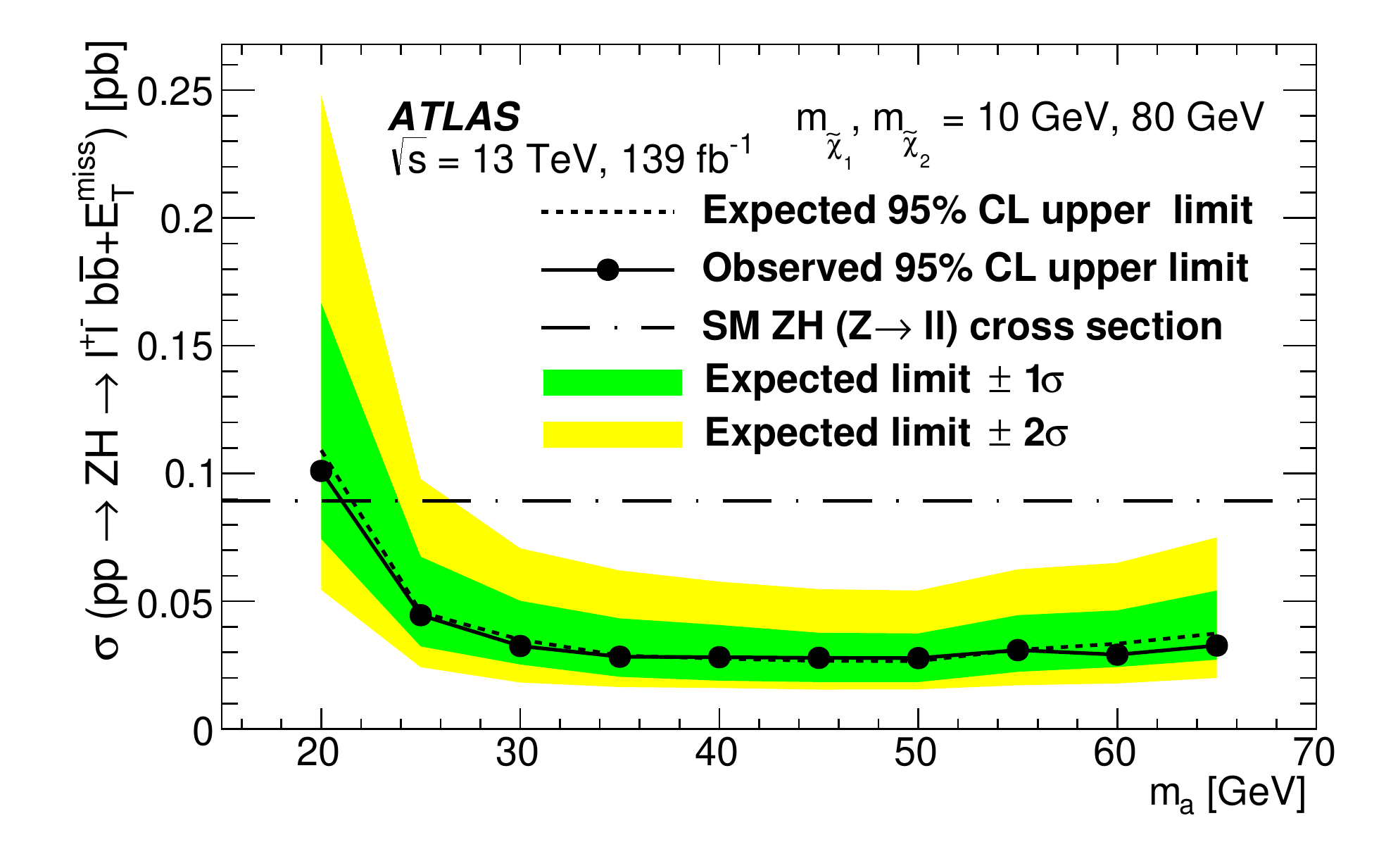}
\includegraphics[width=0.45\textwidth]{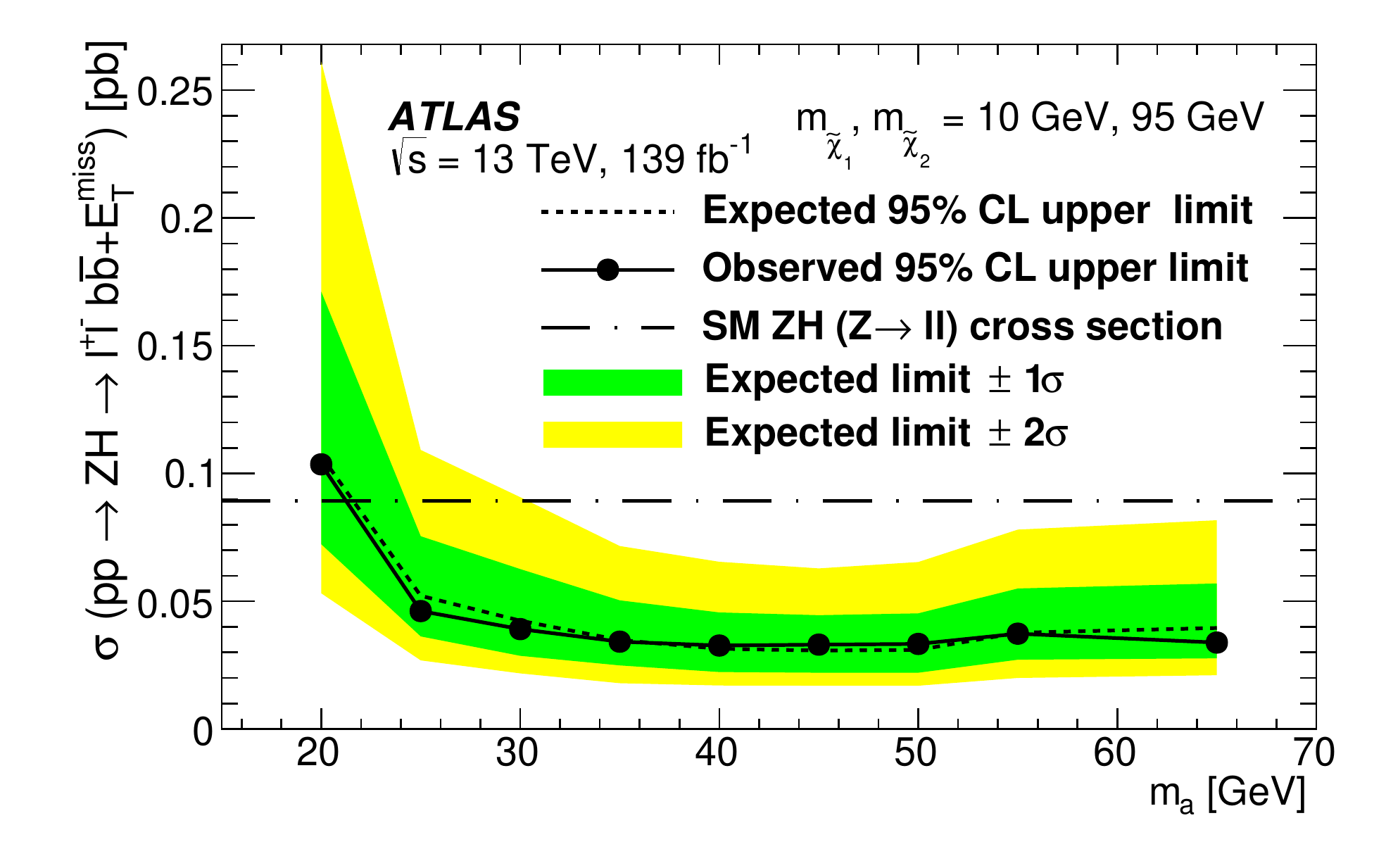}
\includegraphics[width=0.45\textwidth]{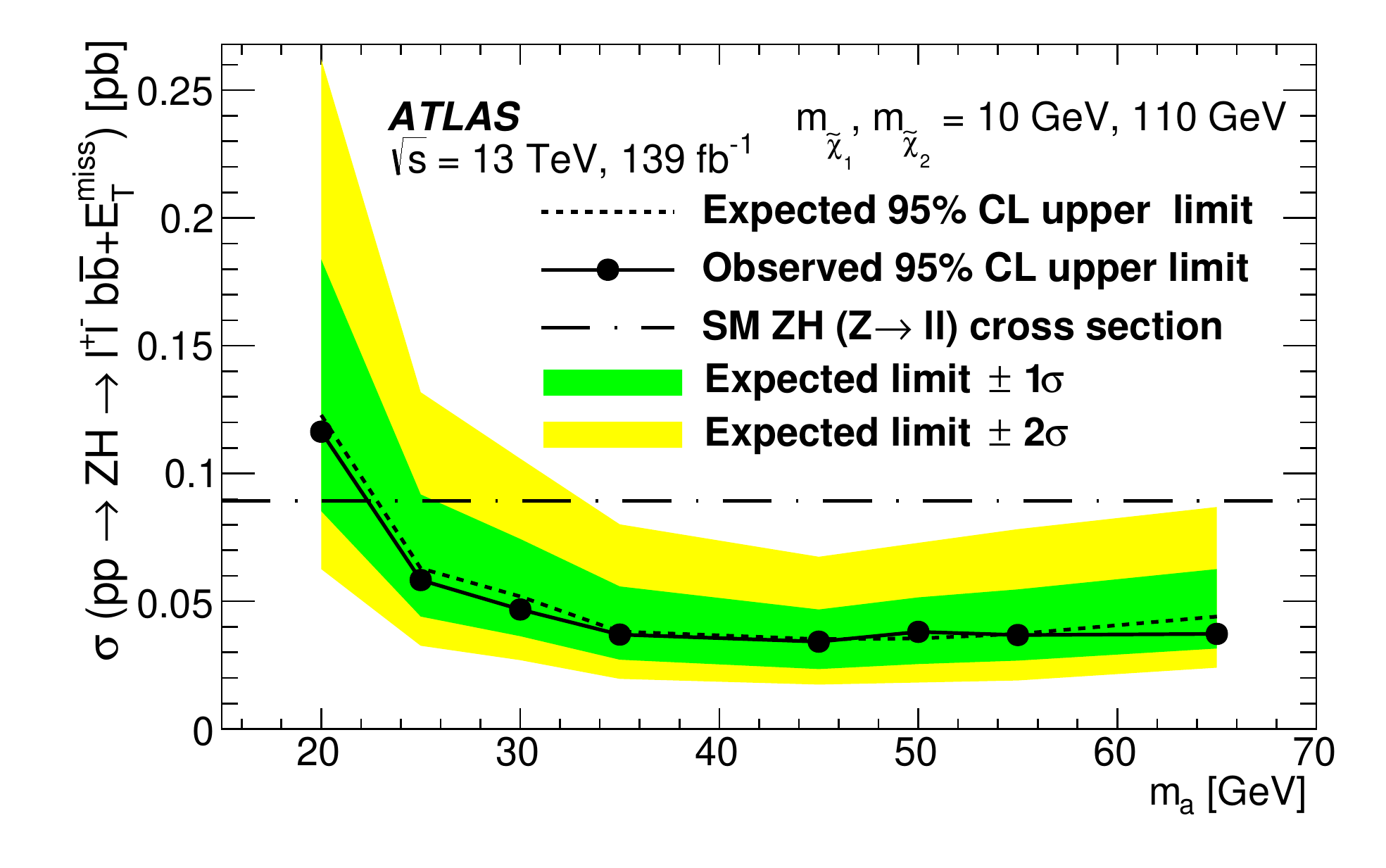}
\includegraphics[width=0.45\textwidth]{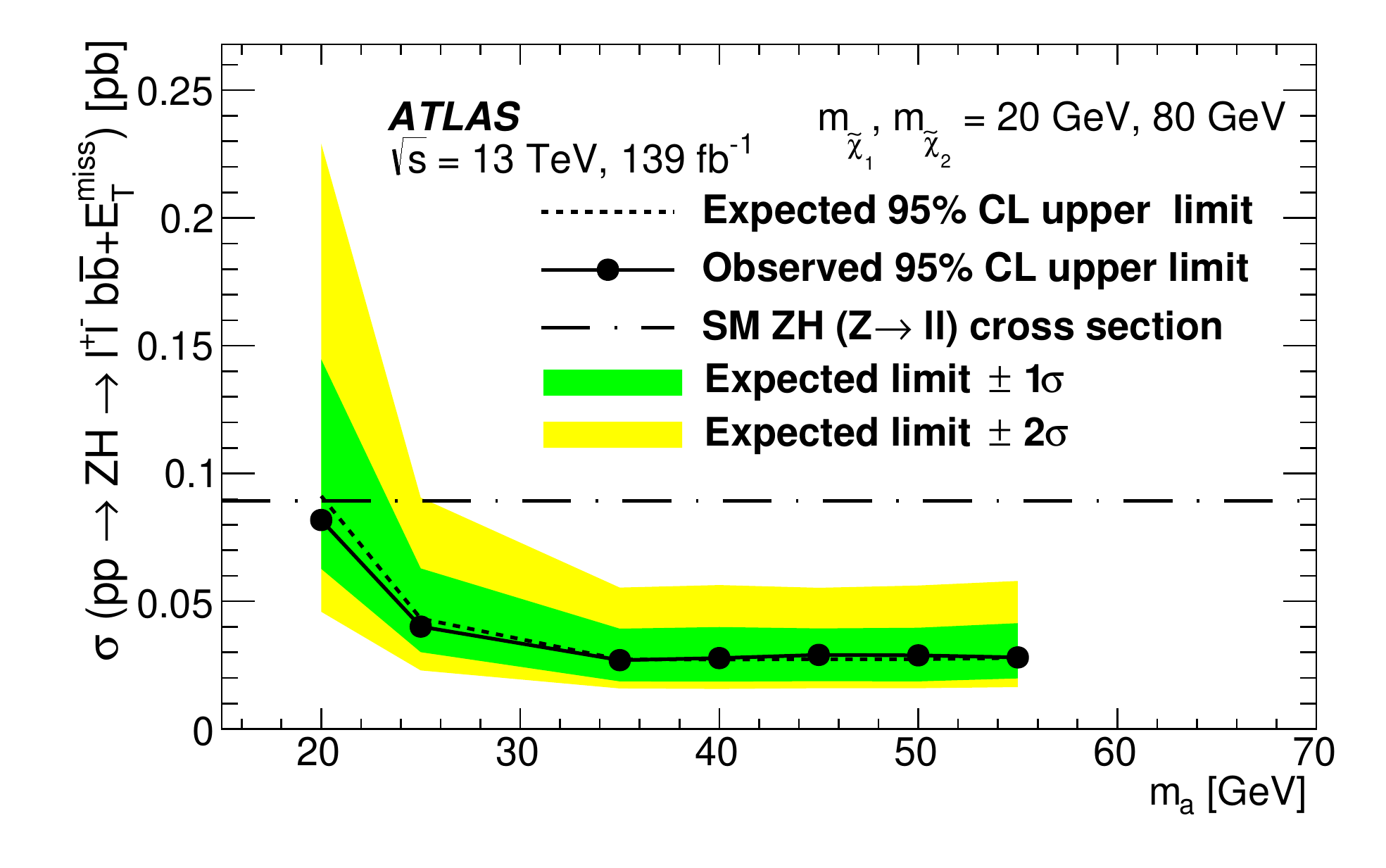}
\includegraphics[width=0.45\textwidth]{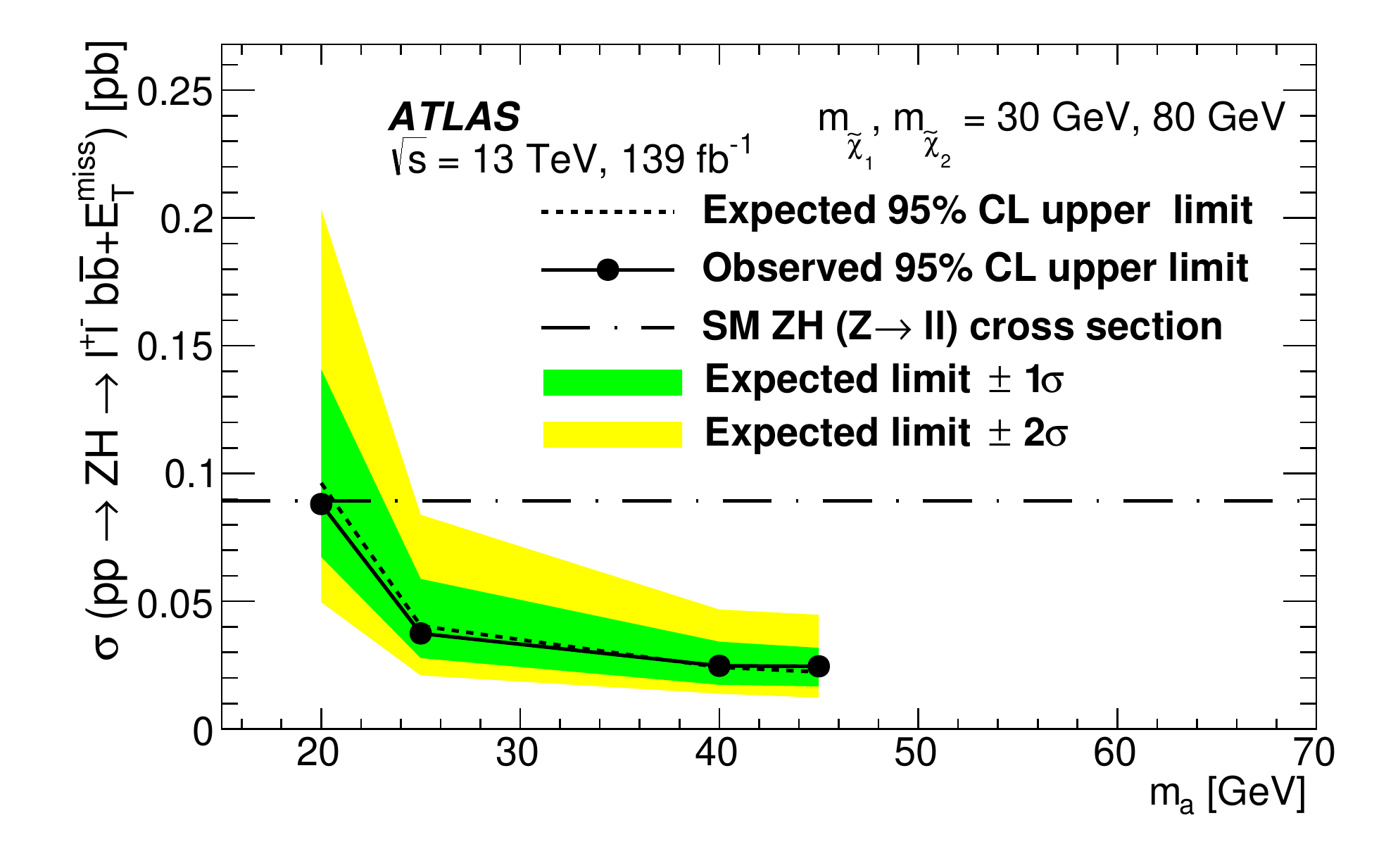}
\caption{Upper limits at 95\% CL on the $pp \rightarrow \ZH$ cross section
times the branching ratio for $Z \rightarrow \ell^{+}\ell^{-}$
(where $\ell = e, \mu~ \mathrm{or}~  \tau$) and $\Hboson \rightarrow
\ninotwo\ninoone \rightarrow a \ninoone\ninoone \rightarrow b\bar{b}
\ninoone\ninoone$  as a
function of \ma for several values of $\mone$ and $\mtwo$ for the
NMSSM scenario described in the text. All branching ratios in the
decay chain after the decay $\Hboson \rightarrow \ninotwo\ninoone$
are set to 100\%.  The
different ranges of \ma points on the horizontal axis reflect
differences in the allowed
event kinematics. The
green and yellow bands show respectively the $\pm 1 \sigma$ and
$\pm 2 \sigma$ ranges for the expected limits.  The lines joining
the \ma points come from an assumed linear interpolation of the
limits. The SM value for the cross section $\sigma(pp \rightarrow
\ZH) \times BR(Z \rightarrow \ell^{+}\ell^{-})$ is shown for reference.}
\label{fig:brazil_plot}
\end{center}
\end{figure}
 
\begin{figure}[htbp]
\begin{center}
\includegraphics[width=0.80\textwidth]{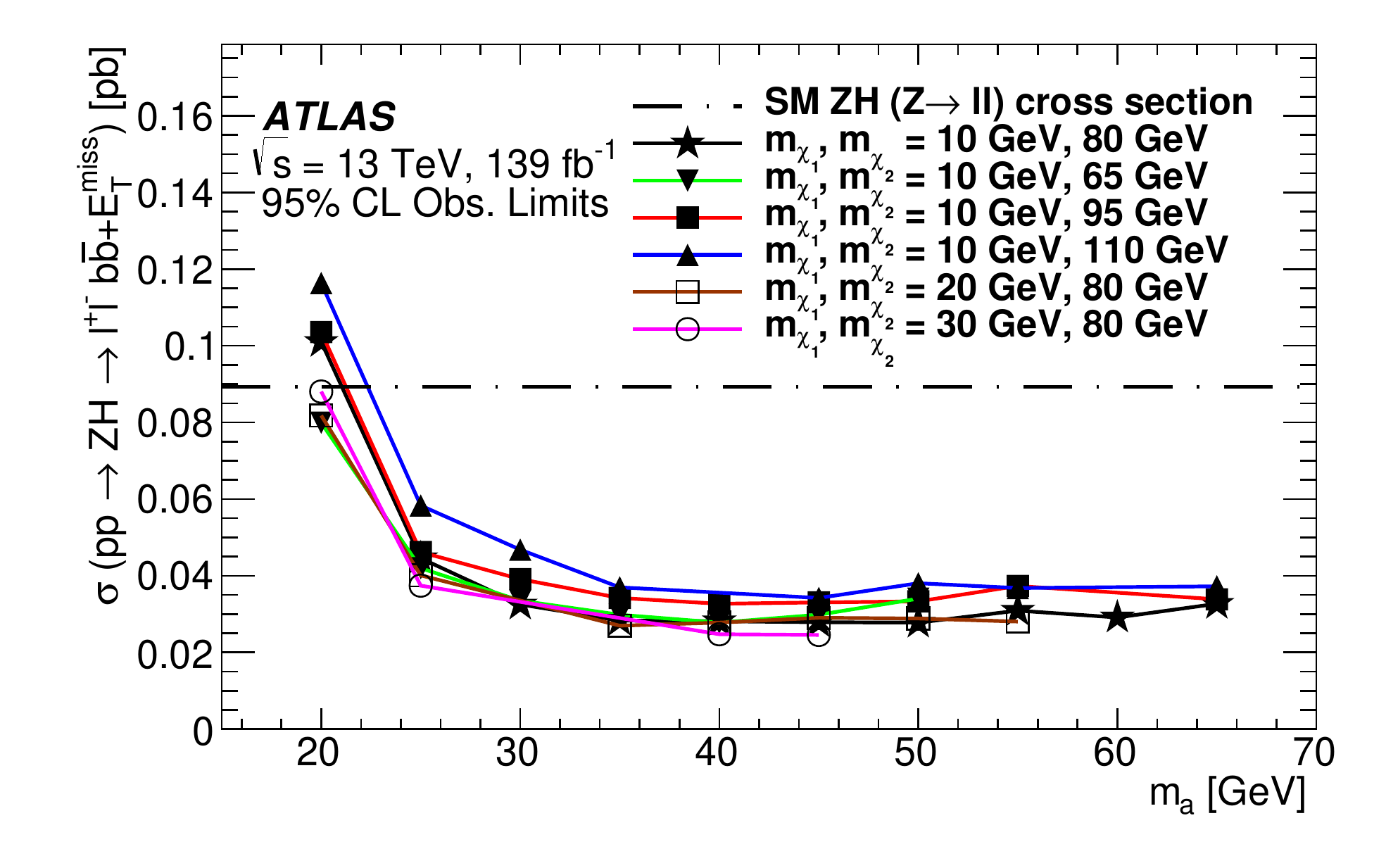}
\caption{
Upper limits at 95\% CL on the cross section $pp \rightarrow \ZH$ times
branching ratio for $Z \rightarrow \ell^{+}\ell^{-}$ (where $\ell =
e, \mu~ \mathrm{or}~ \tau$) and $\Hboson \rightarrow \ninotwo\ninoone
\rightarrow a \ninoone\ninoone \rightarrow b\bar{b} \ninoone\ninoone$  as a
function of \ma for several values of $\mone$ and $\mtwo$ for the
NMSSM scenario described in the text. All branching ratios in the
Higgs boson decay chain after the decay
$\Hboson \rightarrow \ninotwo\ninoone$
are set to 100\%. The
different ranges in \ma reflect
differences in the allowed
event kinematics. The lines joining
the \ma points come from an assumed linear interpolation of the
limits. The SM value for the cross section $\sigma(pp \rightarrow
\ZH) \times BR(Z \rightarrow \ell^{+}\ell^{-})$ is shown for reference.}
\label{fig:UL_all}
\end{center}
\end{figure}
 
\section{Conclusion}
 
A search for the exotic decay of the Higgs boson (\Hboson) into a \bbbar
resonance plus missing transverse momentum
has been performed with the ATLAS detector at the Large Hadron Collider
in \intlumi~\ifb\ of \pp collisions at $\rts = 13$~\TeV\ .
The search was designed to target
Higgs bosons produced in association with a \Zboson boson
and was  conducted in
events with two leptons, two or more jets, at least one of which must be $b$-tagged, and missing transverse
momentum. The analysis was optimized on a model in the Peccei--Quinn
symmetry limit of the NMSSM where
$\Hboson\rightarrow\ninotwo\ninoone$, with
$\ninotwo\rightarrow{a}\ninoone$, and $a$ is a new, light
pseudoscalar Higgs boson.  The decay of the $a$ boson into a pair of
\bquarks results in a resonance in the dijet invariant mass. Such
models differ from those considered in past searches for exotic
Higgs boson decays involving the production of $a$, where the NMSSM
was considered in the $R$-symmetry limit in which the
dominant decay channel is $\Hboson\rightarrow{aa}$.
Observations are consistent with SM expectations and upper limits
on the $pp \rightarrow \ZH$ cross section times the branching ratio
for $Z \rightarrow \ell^{+}\ell^{-}$ and $\Hboson \rightarrow
\ninotwo\ninoone \rightarrow a \ninoone\ninoone \rightarrow b\bar{b}
\ninoone\ninoone$
have been obtained
for a three-dimensional scan of masses of the \ninoone, \ninotwo  and $a$
boson. Assuming the SM cross section for \ZH\ production,
and assuming 100\% branching ratios for the decays $\ninotwo
\rightarrow a \ninoone$ and $a \rightarrow b\overline{b}$, an upper limit
on the branching ratio BR$(\Hboson \rightarrow \ninotwo\ninoone$) of 31\%
is obtained at 95\% confidence level for a range of
$m_{a}$ values between 35 and 55~\GeV\ for fixed values of
$m_{\ninoone} = 10$~\GeV\ and $m_{\ninotwo} = 80$~\GeV.   These
represent the first direct limits on this
exotic Higgs boson decay from the LHC.

\section*{Acknowledgements}
 

We thank CERN for the very successful operation of the LHC, as well as the
support staff from our institutions without whom ATLAS could not be
operated efficiently.
 
We acknowledge the support of
ANPCyT, Argentina;
YerPhI, Armenia;
ARC, Australia;
BMWFW and FWF, Austria;
ANAS, Azerbaijan;
SSTC, Belarus;
CNPq and FAPESP, Brazil;
NSERC, NRC and CFI, Canada;
CERN;
ANID, Chile;
CAS, MOST and NSFC, China;
Minciencias, Colombia;
MSMT CR, MPO CR and VSC CR, Czech Republic;
DNRF and DNSRC, Denmark;
IN2P3-CNRS and CEA-DRF/IRFU, France;
SRNSFG, Georgia;
BMBF, HGF and MPG, Germany;
GSRI, Greece;
RGC and Hong Kong SAR, China;
ISF and Benoziyo Center, Israel;
INFN, Italy;
MEXT and JSPS, Japan;
CNRST, Morocco;
NWO, Netherlands;
RCN, Norway;
MEiN, Poland;
FCT, Portugal;
MNE/IFA, Romania;
JINR;
MES of Russia and NRC KI, Russian Federation;
MESTD, Serbia;
MSSR, Slovakia;
ARRS and MIZ\v{S}, Slovenia;
DSI/NRF, South Africa;
MICINN, Spain;
SRC and Wallenberg Foundation, Sweden;
SERI, SNSF and Cantons of Bern and Geneva, Switzerland;
MOST, Taiwan;
TAEK, Turkey;
STFC, United Kingdom;
DOE and NSF, United States of America.
In addition, individual groups and members have received support from
BCKDF, CANARIE, Compute Canada and CRC, Canada;
COST, ERC, ERDF, Horizon 2020 and Marie Sk{\l}odowska-Curie Actions, European Union;
Investissements d'Avenir Labex, Investissements d'Avenir Idex and ANR, France;
DFG and AvH Foundation, Germany;
Herakleitos, Thales and Aristeia programmes co-financed by EU-ESF and the Greek NSRF, Greece;
BSF-NSF and GIF, Israel;
Norwegian Financial Mechanism 2014-2021, Norway;
NCN and NAWA, Poland;
La Caixa Banking Foundation, CERCA Programme Generalitat de Catalunya and PROMETEO and GenT Programmes Generalitat Valenciana, Spain;
G\"{o}ran Gustafssons Stiftelse, Sweden;
The Royal Society and Leverhulme Trust, United Kingdom.
 
The crucial computing support from all WLCG partners is acknowledged gratefully, in particular from CERN, the ATLAS Tier-1 facilities at TRIUMF (Canada), NDGF (Denmark, Norway, Sweden), CC-IN2P3 (France), KIT/GridKA (Germany), INFN-CNAF (Italy), NL-T1 (Netherlands), PIC (Spain), ASGC (Taiwan), RAL (UK) and BNL (USA), the Tier-2 facilities worldwide and large non-WLCG resource providers. Major contributors of computing resources are listed in Ref.~\cite{ATL-SOFT-PUB-2021-003}.
 

\clearpage
 
\printbibliography

@article{Curtin:2013fra,
      author         = "Curtin, D. and others",
      title          = "{Exotic decays of the 125 GeV Higgs boson}",
      journal        = "Phys. Rev. D",
      number         = "7",
      volume         = "90",
      pages          = "075004",
      doi            = "10.1103/PhysRevD.90.075004",
      year           = "2014",
      eprint         = "1312.4992",
      archivePrefix  = "arXiv",
      primaryClass   = "hep-ph",
      reportNumber   = "YITP-13-47, PITT-PACC-1314",
      SLACcitation   = "%%CITATION = ARXIV:1312.4992;%%",
}

@Article{Fayet:1979sa,
     author    = "Fayet, P.",
     title     = "{Relations between the masses of the superpartners of
                  leptons and quarks, the goldstino coupling and the neutral
                  currents}",
     journal   = "Phys. Lett. B",
     volume    = "84",
     year      = "1979",
     pages     = "416",
     doi       = "10.1016/0370-2693(79)91229-2",
     SLACcitation  = "%%CITATION = PHLTA,B84,416;%%"
}

@article{Huang:2013ima,
      author         = "Huang, Jinrui and Liu, Tao and Wang, Lian-Tao and Yu,
                        Felix",
      title          = "{Supersymmetric Exotic Decays of the 125 GeV Higgs
                        Boson}",
      journal        = "Phys. Rev. Lett.",
      volume         = "112",
      year           = "2014",
      number         = "22",
      pages          = "221803",
      doi            = "10.1103/PhysRevLett.112.221803",
      eprint         = "1309.6633",
      archivePrefix  = "arXiv",
      primaryClass   = "hep-ph",
      reportNumber   = "FERMILAB-PUB-13-186-T, LA-UR-13-24173",
      SLACcitation   = "%%CITATION = ARXIV:1309.6633;%%"
}

@article{Huang:2014cla,
      author         = "Huang, Jinrui and Liu, Tao and Wang, Lian-Tao and Yu,
                        Felix",
      title          = "{Supersymmetric subelectroweak scale dark matter, the
                        Galactic Center gamma-ray excess, and exotic decays of the
                        125 GeV Higgs boson}",
      journal        = "Phys. Rev. D",
      volume         = "90",
      year           = "2014",
      number         = "11",
      pages          = "115006",
      doi            = "10.1103/PhysRevD.90.115006",
      eprint         = "1407.0038",
      archivePrefix  = "arXiv",
      primaryClass   = "hep-ph",
      reportNumber   = "FERMILAB-PUB-14-183-T",
      SLACcitation   = "%%CITATION = ARXIV:1407.0038;%%"
}

@article{Avoni:2018iuv,
  author         = "{G. Avoni et al.}",
      title          = "{The new LUCID-2 detector for luminosity measurement and
                        monitoring in ATLAS}",
      journal        = "JINST",
      volume         = "13",
      year           = "2018",
      number         = "07",
      pages          = "P07017",
      doi            = "10.1088/1748-0221/13/07/P07017",
      SLACcitation   = "%%CITATION = JINST,13,P07017;%%"
}

@article{Ellwanger:2009dp,
    author = "Ellwanger, Ulrich and Hugonie, Cyril and Teixeira, Ana M.",
    title = "{The Next-to-Minimal Supersymmetric Standard Model}",
    eprint = "0910.1785",
    archivePrefix = "arXiv",
    primaryClass = "hep-ph",
    reportNumber = "LPT-ORSAY-09-76, CFTP-09-032, LPTA-09-066",
    doi = "10.1016/j.physrep.2010.07.001",
    journal = "Phys. Rept.",
    volume = "496",
    pages = "1--77",
    year = "2010"
}

@Article{little_hierarchy,
    author = "Barbieri, Riccardo and Strumia, Alessandro",
    title = "{The 'LEP paradox'}",
    year = "2000",
    booktitle = "{4th Rencontres du Vietnam}: {Physics at Extreme Energies (Particle Physics and Astrophysics)}",
    eprint = "hep-ph/0007265",
    archivePrefix = "arXiv",
    reportNumber = "IFUP-TH-2000-22, SNS-PH-00-12",
    month = "7",
  
}

@article{Fayet:1976et,
      author         = "Fayet, Pierre",
      title          = "{Supersymmetry and weak, electromagnetic and strong
                        interactions}",
      journal        = "Phys. Lett. B",
      volume         = "64",
      year           = "1976",
      pages          = "159",
      doi            = "10.1016/0370-2693(76)90319-1",
      reportNumber   = "PTENS 76/14",
      SLACcitation   = "%%CITATION = PHLTA,B64,159;%%"
}

@article{Fayet:1977yc,
      author         = "Fayet, Pierre",
      title          = "{Spontaneously broken supersymmetric theories of weak,
                        electromagnetic and strong interactions}",
      journal        = "Phys. Lett. B",
      volume         = "69",
      year           = "1977",
      pages          = "489",
      doi            = "10.1016/0370-2693(77)90852-8",
      reportNumber   = "LPTENS 77/11",
      SLACcitation   = "%%CITATION = PHLTA,B69,489;%%"
}

@article{Farrar:1978xj,
      author         = "Farrar, Glennys R. and Fayet, Pierre",
      title          = "{Phenomenology of the production, decay, and detection of
                        new hadronic states associated with supersymmetry}",
      journal        = "Phys. Lett. B",
      volume         = "76",
      year           = "1978",
      pages          = "575",
      doi            = "10.1016/0370-2693(78)90858-4",
      reportNumber   = "CALT-68-648",
      SLACcitation   = "%%CITATION = PHLTA,B76,575;%%"
}

@article{Dimopoulos:1981zb,
      author         = "Dimopoulos, Savas and Georgi, Howard",
      title          = "{Softly broken supersymmetry and SU(5)}",
      journal        = "Nucl. Phys. B",
      volume         = "193",
      year           = "1981",
      pages          = "150",
      doi            = "10.1016/0550-3213(81)90522-8",
      reportNumber   = "HUTP-81/A022",
      SLACcitation   = "%%CITATION = NUPHA,B193,150;%%"
}

@Article{Evans:2008zzb,
      author         = "Evans, Lyndon and Bryant, Philip",
      title          = "{LHC Machine}",
      journal        = "JINST",
      volume         = "3",
      pages          = "S08001",
      doi            = "10.1088/1748-0221/3/08/S08001",
      year           = "2008",
      SLACcitation   = "%%CITATION = JINST,3,S08001;%%",
}

@Article{Cacciari:2008gp,
     author    = "Cacciari, Matteo and Salam, Gavin P. and Soyez, Gregory",
     title     = "{The anti-\(k_{t}\) jet clustering algorithm}",
     journal   = "JHEP",
     volume    = "04",
     year      = "2008",
     pages     = "063",
     eprint    = "0802.1189",
     archivePrefix = "arXiv",
     primaryClass  =  "hep-ph",
     doi       = "10.1088/1126-6708/2008/04/063",
     SLACcitation  = "%%CITATION = 0802.1189;%%"
}

@Article{Fastjet,
      author         = "Cacciari, Matteo and Salam, Gavin P. and Soyez, Gregory",
      title          = "{FastJet user manual}",
      journal        = "Eur. Phys. J. C",
      volume         = "72",
      year           = "2012",
      pages          = "1896",
      doi            = "10.1140/epjc/s10052-012-1896-2",
      eprint         = "1111.6097",
      archivePrefix  = "arXiv",
      primaryClass   = "hep-ph",
      reportNumber   = "CERN-PH-TH-2011-297",
      SLACcitation   = "%%CITATION = ARXIV:1111.6097;%%"
}

@Article{Pumplin:2002vw,
     author    = "Pumplin, J. and others",
     title     = "{New Generation of Parton Distributions with Uncertainties from Global QCD Analysis}",
     journal   = "JHEP",
     volume    = "07",
     year      = "2002",
     pages     = "012",
     doi       = "10.1088/1126-6708/2002/07/012",
     eprint    = "hep-ph/0201195",
     archivePrefix = "arXiv",
     SLACcitation  = "%%CITATION = HEP-PH/0201195;%%"
}

@article{Ball:2012cx,
      author         = "Ball, Richard D. and others",
      title          = "{Parton distributions with LHC data}",
      journal        = "Nucl. Phys. B",
      volume         = "867",
      year           = "2013",
      pages          = "244",
      doi            = "10.1016/j.nuclphysb.2012.10.003",
      eprint         = "1207.1303",
      archivePrefix  = "arXiv",
      primaryClass   = "hep-ph",
      reportNumber   = "EDINBURGH-2012-08, IFUM-FT-997, FR-PHENO-2012-014,
                        RWTH-TTK-12-25, CERN-PH-TH-2012-037, SFB-CPP-12-47\,
                        --CERN-PH-TH-2012-037",
      SLACcitation   = "%%CITATION = ARXIV:1207.1303;%%"
}

@article{Ball:2014uwa,
      author         = "Ball, Richard D. and others",
      title          = "{Parton distributions for the LHC run II}",
      collaboration  = "NNPDF",
      journal        = "JHEP",
      volume         = "04",
      year           = "2015",
      pages          = "040",
      doi            = "10.1007/JHEP04(2015)040",
      eprint         = "1410.8849",
      archivePrefix  = "arXiv",
      primaryClass   = "hep-ph",
      reportNumber   = "EDINBURGH-2014-15, IFUM-1034-FT, CERN-PH-TH-2013-253,
                        OUTP-14-11P, CAVENDISH-HEP-14-11",
      SLACcitation   = "%%CITATION = ARXIV:1410.8849;%%"
}

@article{Czakon:2012pz,
      author         = "Czakon, Michal and Mitov, Alexander",
      title          = "{NNLO corrections to top pair production at hadron
                        colliders: the quark-gluon reaction}",
      journal        = "JHEP",
      volume         = "01",
      pages          = "080",
      doi            = "10.1007/JHEP01(2013)080",
      year           = "2013",
      eprint         = "1210.6832",
      archivePrefix  = "arXiv",
      primaryClass   = "hep-ph",
      SLACcitation   = "%%CITATION = ARXIV:1210.6832;%%",
}

@Article{Sjostrand:2007gs,
	author    = "Sj{\"o}strand, T. and Mrenna, S. and Skands, P.",
	title     = "{A brief introduction to PYTHIA 8.1}",
	journal   = "Comput. Phys. Commun.",
	volume    = "178",
	year      = "2008",
	pages     = "852-867",
	eprint    = "0710.3820",
	archivePrefix = "arXiv",
	primaryClass  =  "hep-ph",
	doi       = "10.1016/j.cpc.2008.01.036",
	SLACcitation  = "%%CITATION = 0710.3820;%%"
}

@article{Sjostrand:2014zea,
      author         = "Sj{\"o}strand, Torbj{\"o}rn and Ask, Stefan and Christiansen,
                        Jesper R. and Corke, Richard and Desai, Nishita and Ilten,
                        Philip and Mrenna, Stephen and Prestel, Stefan and
                        Rasmussen, Christine O. and Skands, Peter Z.",
      title          = "{An introduction to PYTHIA 8.2}",
      journal        = "Comput. Phys. Commun.",
      volume         = "191",
      year           = "2015",
      pages          = "159",
      doi            = "10.1016/j.cpc.2015.01.024",
      eprint         = "1410.3012",
      archivePrefix  = "arXiv",
      primaryClass   = "hep-ph",
      reportNumber   = "LU-TP-14-36, MCNET-14-22, CERN-PH-TH-2014-190,
                        FERMILAB-PUB-14-316-CD, DESY-14-178, SLAC-PUB-16122,
                        --FERMILAB-PUB-14-316-CD",
      SLACcitation   = "%%CITATION = ARXIV:1410.3012;%%"
}

@Article{Alwall:2014hca,
      author         = "Alwall, J. and Frederix, R. and Frixione, S. and Hirschi,
                        V. and Maltoni, F. and Mattelaer, O. and Shao, H. -S. and
                        Stelzer, T. and Torrielli, P. and Zaro, M.",
      title          = "{The automated computation of tree-level and
                        next-to-leading order differential cross sections, and
                        their matching to parton shower simulations}",
      journal        = "JHEP",
      volume         = "07",
      year           = "2014",
      pages          = "079",
      doi            = "10.1007/JHEP07(2014)079",
      eprint         = "1405.0301",
      archivePrefix  = "arXiv",
      primaryClass   = "hep-ph",
      reportNumber   = "CERN-PH-TH-2014-064, CP3-14-18, LPN14-066, MCNET-14-09,
                        ZU-TH-14-14",
      SLACcitation   = "%%CITATION = ARXIV:1405.0301;%%"
}

@Article{Bahr:2008pv,
      author         = "B{\"a}hr, M. and others",
      title          = "{Herwig++ physics and manual}",
      journal        = "Eur. Phys. J. C",
      volume         = "58",
      year           = "2008",
      pages          = "639",
      doi            = "10.1140/epjc/s10052-008-0798-9",
      eprint         = "0803.0883",
      archivePrefix  = "arXiv",
      primaryClass   = "hep-ph",
      reportNumber   = "CERN-PH-TH-2008-038, CAVENDISH-HEP-08-03, KA-TP-05-2008,
                        DCPT-08-22, IPPP-08-11, CP3-08-05",
      SLACcitation   = "%%CITATION = ARXIV:0803.0883;%%"
}

@Article{Bellm:2015jjp,
      author         = "Bellm, Johannes and others",
      title          = "{Herwig 7.0/Herwig++ 3.0 release note}",
      journal        = "Eur. Phys. J. C",
      volume         = "76",
      year           = "2016",
      number         = "4",
      pages          = "196",
      doi            = "10.1140/epjc/s10052-016-4018-8",
      eprint         = "1512.01178",
      archivePrefix  = "arXiv",
      primaryClass   = "hep-ph",
      reportNumber   = "CERN-PH-TH-2015-289, MAN-HEP-2015-15, IFJPAN-IV-2015-13,
                        HERWIG-2015-01, KA-TP-18-2015, DCPT-15-142, MCNET-15-28,
                        IPPP-15-71, --HERWIG-2015-01",
      SLACcitation   = "%%CITATION = ARXIV:1512.01178;%%"
}

@Article{Nason:2004rx,
      author         = "Nason, Paolo",
      title          = "{A new method for combining NLO QCD with shower Monte Carlo algorithms}",
      journal        = "JHEP",
      volume         = "11",
      pages          = "040",
      doi            = "10.1088/1126-6708/2004/11/040",
      year           = "2004",
      eprint         = "hep-ph/0409146",
      archivePrefix  = "arXiv",
}

@Article{Frixione:2007vw,
      author         = "Frixione, Stefano and Nason, Paolo and Oleari, Carlo",
      title          = "{Matching NLO QCD computations with parton shower
                        simulations: the POWHEG method}",
      journal        = "JHEP",
      volume         = "11",
      pages          = "070",
      doi            = "10.1088/1126-6708/2007/11/070",
      year           = "2007",
      eprint         = "0709.2092",
      archivePrefix  = "arXiv",
      primaryClass   = "hep-ph",
}

@Article{Alioli:2010xd,
      author         = "Alioli, Simone and Nason, Paolo and Oleari, Carlo and Re,
                        Emanuele",
      title          = "{A general framework for implementing NLO calculations in
                        shower Monte Carlo programs: the POWHEG BOX}",
      journal        = "JHEP",
      volume         = "06",
      pages          = "043",
      doi            = "10.1007/JHEP06(2010)043",
      year           = "2010",
      eprint         = "1002.2581",
      archivePrefix  = "arXiv",
      primaryClass   = "hep-ph",
}

@article{Re:2010bp,
      author         = "Re, Emanuele",
      title          = "{Single-top \(Wt\)-channel production matched with parton
                        showers using the POWHEG method}",
      journal        = "Eur. Phys. J. C",
      volume         = "71",
      year           = "2011",
      pages          = "1547",
      doi            = "10.1140/epjc/s10052-011-1547-z",
      eprint         = "1009.2450",
      archivePrefix  = "arXiv",
      primaryClass   = "hep-ph",
      reportNumber   = "IPPP-10-74, DCPT-10-148",
      SLACcitation   = "%%CITATION = ARXIV:1009.2450;%%"
}

@article{Kidonakis:2010ux,
      author         = "Kidonakis, Nikolaos",
      title          = "{Two-loop soft anomalous dimensions for single top quark
                        associated production with a \(W^{-}\) or \(H^{-}\)}",
      journal        = "Phys. Rev. D",
      volume         = "82",
      year           = "2010",
      pages          = "054018",
      doi            = "10.1103/PhysRevD.82.054018",
      eprint         = "1005.4451",
      archivePrefix  = "arXiv",
      primaryClass   = "hep-ph",
      SLACcitation   = "%%CITATION = ARXIV:1005.4451;%%"
}

@article{Beneke:2011mq,
      author         = "Beneke, M. and Falgari, P. and Klein, S. and Schwinn, C.",
      title          = "{Hadronic top-quark pair production with NNLL threshold
                        resummation}",
      journal        = "Nucl. Phys. B",
      volume         = "855",
      year           = "2012",
      pages          = "695-741",
      doi            = "10.1016/j.nuclphysb.2011.10.021",
      eprint         = "1109.1536",
      archivePrefix  = "arXiv",
      primaryClass   = "hep-ph",
      reportNumber   = "TTK-11-38, ITP-UU-11-26, SPIN-11-19, FR-PHENO-2011-015,
                        SFB-CPP-11-49",
      SLACcitation   = "%%CITATION = ARXIV:1109.1536;%%"
}

@article{Anastasiou:2003ds,
      author         = "Anastasiou, Charalampos and Dixon, Lance J. and Melnikov,
                        Kirill and Petriello, Frank",
      title          = "{High precision QCD at hadron colliders: Electroweak
                        gauge boson rapidity distributions at next-to-next-to leading order}",
      journal        = "Phys. Rev. D",
      volume         = "69",
      year           = "2004",
      pages          = "094008",
      doi            = "10.1103/PhysRevD.69.094008",
      eprint         = "hep-ph/0312266",
      archivePrefix  = "arXiv",
      reportNumber   = "SLAC-PUB-10288, UH-511-1042-03",
      SLACcitation   = "%%CITATION = HEP-PH/0312266;%%"
}

@article{Cacciari:2011hy,
      author         = "Cacciari, Matteo and Czakon, Michal and Mangano,
                        Michelangelo and Mitov, Alexander and Nason, Paolo",
      title          = "{Top-pair production at hadron colliders with
                        next-to-next-to-leading logarithmic soft-gluon
                        resummation}",
      journal        = "Phys. Lett. B",
      volume         = "710",
      pages          = "612-622",
      doi            = "10.1016/j.physletb.2012.03.013",
      year           = "2012",
      eprint         = "1111.5869",
      primaryClass   = "hep-ph",
      archivePrefix  = "arXiv",
      reportNumber   = "CERN-PH-TH-2011-277, TTK-11-54",
      SLACcitation   = "%%CITATION = ARXIV:1111.5869;%%",
}

@article{Czakon:2012zr,
      author         = "Czakon, Michal and Mitov, Alexander",
      title          = "{NNLO corrections to top-pair production at hadron
                        colliders: the all-fermionic scattering channels}",
      journal        = "JHEP",
      volume         = "12",
      pages          = "054",
      doi            = "10.1007/JHEP12(2012)054",
      year           = "2012",
      eprint         = "1207.0236",
      archivePrefix  = "arXiv",
      primaryClass   = "hep-ph",
      SLACcitation   = "%%CITATION = ARXIV:1207.0236;%%",
}

@article{Czakon:2013goa,
      author         = "Czakon, Michal and Fiedler, Paul and Mitov, Alexander",
      title          = "{Total Top-Quark Pair-Production Cross Section at
                        Hadron Colliders Through \(O(\alpha_S^4)\)}",
      journal        = "Phys. Rev. Lett.",
      volume         = "110",
      pages          = "252004",
      doi            = "10.1103/PhysRevLett.110.252004",
      year           = "2013",
      eprint         = "1303.6254",
      archivePrefix  = "arXiv",
      primaryClass   = "hep-ph",
      reportNumber   = "CERN-PH-TH-2013-056, TTK-13-08",
      SLACcitation   = "%%CITATION = ARXIV:1303.6254;%%",
}

@article{Baernreuther:2012ws,
      author         = "B{\"a}rnreuther, Peter and Czakon, Michal and Mitov,
                        Alexander",
      title          = "{Percent-Level-Precision Physics at the
                  Tevatron: Next-to-Next-to-Leading Order QCD
                  Corrections to \(q \bar{q} \to t \bar{t} + X\)}",
      journal        = "Phys. Rev. Lett.",
      volume         = "109",
      pages          = "132001",
      doi            = "10.1103/PhysRevLett.109.132001",
      year           = "2012",
      eprint         = "1204.5201",
      archivePrefix  = "arXiv",
      primaryClass   = "hep-ph",
      SLACcitation   = "%%CITATION = ARXIV:1204.5201;%%",
}

@article{Frixione:2007nw,
      author         = "Frixione, Stefano and Nason, Paolo and Ridolfi, Giovanni",
      title          = "{A positive-weight next-to-leading-order Monte Carlo for
                        heavy flavour hadroproduction}",
      journal        = "JHEP",
      volume         = "09",
      pages          = "126",
      doi            = "10.1088/1126-6708/2007/09/126",
      year           = "2007",
      eprint         = "0707.3088",
      archivePrefix  = "arXiv",
      primaryClass   = "hep-ph",
      SLACcitation   = "%%CITATION = ARXIV:0707.3088;%%",
}

@article{deFlorian:2016spz,
      author         = "de Florian, D. and others",
      title          = "{Handbook of LHC Higgs Cross Sections: 4. Deciphering the
                        Nature of the Higgs Sector}",
      collaboration  = "LHC Higgs Cross Section Working Group",
      doi            = "10.23731/CYRM-2017-002",
      year           = "2016",
      eprint         = "1610.07922",
      archivePrefix  = "arXiv",
      primaryClass   = "hep-ph",
      reportNumber   = "FERMILAB-FN-1025-T, CERN-2017-002-M",
      SLACcitation   = "%%CITATION = ARXIV:1610.07922;%%"
}

@Article{Hoeche:2009rj,
      author         = "H{\"o}che, Stefan and Krauss, Frank and Schumann, Steffen and Siegert, Frank",
      title          = "{QCD matrix elements and truncated showers}",
      journal        = "JHEP",
      volume         = "05",
      pages          = "053",
      doi            = "10.1088/1126-6708/2009/05/053",
      year           = "2009",
      eprint         = "0903.1219",
      archivePrefix  = "arXiv",
      primaryClass   = "hep-ph",
}

@article{Hoeche:2012yf,
      author         = "H{\"o}che, Stefan and Krauss, Frank and Sch{\"o}nherr, Marek and
                        Siegert, Frank",
      title          = "{QCD matrix elements + parton showers. The NLO case}",
      journal        = "JHEP",
      volume         = "04",
      year           = "2013",
      pages          = "027",
      doi            = "10.1007/JHEP04(2013)027",
      eprint         = "1207.5030",
      archivePrefix  = "arXiv",
      primaryClass   = "hep-ph",
      reportNumber   = "SLAC-PUB-15191, IPPP-12-52, DCPT-12-104, LPN12-081,
                        FR-PHENO-2012-017, MCNET-12-09, --FR-PHENO-2012-017",
      SLACcitation   = "%%CITATION = ARXIV:1207.5030;%%"
}

@article{Bothmann:2016nao,
      author         = "Bothmann, Enrico and Sch{\"o}nherr, Marek and Schumann, Steffen",
      title          = "{Reweighting QCD matrix-element and parton-shower calculations}",
      journal        = "Eur. Phys. J. C",
      volume         = "76",
      year           = "2016",
      number         = "11",
      pages          = "590",
      doi            = "10.1140/epjc/s10052-016-4430-0",
      eprint         = "1606.08753",
      archivePrefix  = "arXiv",
      primaryClass   = "hep-ph",
      reportNumber   = "MCNET-16-22, ZU-TH-21-16, ZU--TH--21-16",
      SLACcitation   = "%%CITATION = ARXIV:1606.08753;%%"
}

@article{Bothmann:2019yzt,
      author = "Bothmann, Enrico and others",
      title = "{Event generation with Sherpa 2.2}",
      journal = "SciPost Phys.",
      volume = "7",
      year = "2019",
      number = "3",
      pages = "034",
      doi = "10.21468/SciPostPhys.7.3.034",
      reportNumber = "FERMILAB-PUB-19-218-T, SLAC-PUB-17433, IPPP/19/42, MCNET-19-11",
      eprint = "1905.09127",
      archivePrefix = "arXiv",
      primaryClass = "hep-ph",
}

@article{Hoeche:2011fd,
      author         = "H{\"o}che, Stefan and Krauss, Frank and Sch{\"o}nherr, Marek and
                        Siegert, Frank",
      title          = "{A critical appraisal of NLO+PS matching methods}",
      journal        = "JHEP",
      volume         = "09",
      year           = "2012",
      pages          = "049",
      doi            = "10.1007/JHEP09(2012)049",
      eprint         = "1111.1220",
      archivePrefix  = "arXiv",
      primaryClass   = "hep-ph",
      reportNumber   = "SLAC-PUB-14661, IPPP-11-67, DCPT-11-134, LPN11-58,
                        FR-PHENO-2011-019, MCNET-11-24",
      SLACcitation   = "%%CITATION = ARXIV:1111.1220;%%"
}

@article{Catani:2001cc,
      author         = "Catani, S. and Krauss, F. and Webber, B. R. and Kuhn, R. ",
      title          = "{QCD Matrix Elements + Parton Showers}",
      journal        = "JHEP",
      volume         = "11",
      year           = "2001",
      pages          = "063",
      doi            = "10.1088/1126-6708/2001/11/063",
      eprint         = "hep-ph/0109231",
      archivePrefix  = "arXiv",
      reportNumber   = "CERN-TH-2000-367, CAVENDISH-HEP-00-03",
      SLACcitation   = "%%CITATION = HEP-PH/0109231;%%"
}

@article{Czakon:2011xx,
  author         = "Czakon, Michal and Mitov, Alexander",
  title          = "{Top++: A program for the calculation of the top-pair cross-section at hadron colliders}",
  journal        = "Comput. Phys. Commun.",
  volume         = "185",
  year           = "2014",
  pages          = "2930",
  doi            = "10.1016/j.cpc.2014.06.021",
  eprint         = "1112.5675",
  archivePrefix  = "arXiv",
  primaryClass   = "hep-ph",
  reportNumber   = "CERN-PH-TH-2011-303, TTK-11-58",
}

@article{Agostinelli:2002hh,
      author         = "{GEANT4 Collaboration} and Agostinelli, S. and others",
      title          = "{\textsc{Geant4} -- a simulation toolkit}",
      journal        = "Nucl. Instrum. Meth. A",
      volume         = "506",
      year           = "2003",
      pages          = "250",
      doi            = "10.1016/S0168-9002(03)01368-8",
      reportNumber   = "SLAC-PUB-9350, FERMILAB-PUB-03-339",
      SLACcitation   = "%%CITATION = NUIMA,A506,250;%%"
}

@Article{Read:2002hq,
    Author = {Read, Alexander L.},
    Journal = {J. Phys. G},
    Pages = {2693},
    doi = "10.1088/0954-3899/28/10/313",
    Title = "{Presentation of search results: the \(CL_S\) technique}",
    Volume = {28},
    Year = {2002}
}

@Article{Cowan:2010js,
   Author = {Cowan, Glen and Cranmer, Kyle and Gross, Eilam and Vitells, Ofer},
   Title = {{Asymptotic formulae for likelihood-based tests of new physics}},
   Journal = {Eur. Phys. J. C},
   Volume = {71},
   Year = {2011},
   Pages = {1554},
   doi = "10.1140/epjc/s10052-011-1554-0",
   Eprint = "1007.1727",
   Archiveprefix = {arXiv},
   Primaryclass = {physics.data-an},
   related = "Cowan:2010js-err",
   relatedstring  = "Erratum:",
}

@Article{PERF-2007-01,
    author         = "{ATLAS Collaboration}",
    title          = "{The ATLAS Experiment at the CERN Large Hadron Collider}",
    journal        = "JINST",
    volume         = "3",
    year           = "2008",
    pages          = "S08003",
    doi            = "10.1088/1748-0221/3/08/S08003",
    primaryClass   = "hep-ex",
}

@Article{SOFT-2010-01,
    author         = "{ATLAS Collaboration}",
    title          = "{The ATLAS Simulation Infrastructure}",
    journal        = "Eur. Phys. J. C",
    volume         = "70",
    year           = "2010",
    pages          = "823",
    doi            = "10.1140/epjc/s10052-010-1429-9",
    eprint         = "1005.4568",
    archivePrefix  = "arXiv",
    primaryClass   = "physics.ins-det",
}

@Article{HIGG-2012-27,
    author         = "{ATLAS Collaboration}",
    title          = "{Observation of a new particle in the search for the Standard Model Higgs boson with the ATLAS detector at the LHC}",
    journal        = "Phys. Lett. B",
    volume         = "716",
    year           = "2012",
    pages          = "1",
    doi            = "10.1016/j.physletb.2012.08.020",
    reportNumber   = "CERN-PH-EP-2012-218",
    eprint         = "1207.7214",
    archivePrefix  = "arXiv",
    primaryClass   = "hep-ex",
}

@Article{STDM-2012-23,
    author         = "{ATLAS Collaboration}",
    title          = "{Measurement of the \(Z/\gamma^*\) boson transverse momentum distribution in \(pp\) collisions at \(\sqrt{s} = 7\,\text{TeV}\) with the ATLAS detector}",
    journal        = "JHEP",
    volume         = "09",
    year           = "2014",
    pages          = "145",
    doi            = "10.1007/JHEP09(2014)145",
    reportNumber   = "CERN-PH-EP-2014-075",
    eprint         = "1406.3660",
    archivePrefix  = "arXiv",
    primaryClass   = "hep-ex",
}

@Article{HIGG-2014-02,
    author         = "{ATLAS Collaboration}",
    title          = "{Search for Higgs bosons decaying to \(aa\) in the \(\mu\mu\tau\tau\) final state in \(pp\) collisions at \(\sqrt{s} = 8\,\text{TeV}\) with the ATLAS experiment}",
    journal        = "Phys. Rev. D",
    volume         = "92",
    year           = "2015",
    pages          = "052002",
    doi            = "10.1103/PhysRevD.92.052002",
    reportNumber   = "CERN-PH-EP-2015-057",
    eprint         = "1505.01609",
    archivePrefix  = "arXiv",
    primaryClass   = "hep-ex",
}

@Article{PERF-2014-03,
    author         = "{ATLAS Collaboration}",
    title          = "{Performance of pile-up mitigation techniques for jets in \(pp\) collisions at \(\sqrt{s} = 8\,\text{TeV}\) using the ATLAS detector}",
    journal        = "Eur. Phys. J. C",
    volume         = "76",
    year           = "2016",
    pages          = "581",
    doi            = "10.1140/epjc/s10052-016-4395-z",
    reportNumber   = "CERN-PH-EP-2015-206",
    eprint         = "1510.03823",
    archivePrefix  = "arXiv",
    primaryClass   = "hep-ex",
}

@Article{PERF-2014-07,
    author         = "{ATLAS Collaboration}",
    title          = "{Topological cell clustering in the ATLAS calorimeters and its performance in LHC Run~1}",
    journal        = "Eur. Phys. J. C",
    volume         = "77",
    year           = "2017",
    pages          = "490",
    doi            = "10.1140/epjc/s10052-017-5004-5",
    reportNumber   = "CERN-PH-EP-2015-304",
    eprint         = "1603.02934",
    archivePrefix  = "arXiv",
    primaryClass   = "hep-ex",
}

@Article{HIGG-2015-07,
    author         = "{ATLAS and CMS Collaborations}",
    title          = "{Measurements of the Higgs boson production and decay rates and constraints on its couplings from a combined ATLAS and CMS analysis of the LHC \(pp\) collision data at \(\sqrt{s} = 7\) and \(8\,\text{TeV}\)}",
    journal        = "JHEP",
    volume         = "08",
    year           = "2016",
    pages          = "045",
    doi            = "10.1007/JHEP08(2016)045",
    eprint         = "1606.02266",
    archivePrefix  = "arXiv",
    primaryClass   = "hep-ex",
}

@Article{PERF-2015-08,
    author         = "{ATLAS Collaboration}",
    title          = "{Performance of the ATLAS track reconstruction algorithms in dense environments in LHC Run~2}",
    journal        = "Eur. Phys. J. C",
    volume         = "77",
    year           = "2017",
    pages          = "673",
    doi            = "10.1140/epjc/s10052-017-5225-7",
    reportNumber   = "CERN-EP-2017-045",
    eprint         = "1704.07983",
    archivePrefix  = "arXiv",
    primaryClass   = "hep-ex",
}

@Article{PERF-2015-10,
    author         = "{ATLAS Collaboration}",
    title          = "{Muon reconstruction performance of the ATLAS detector in proton--proton collision data at \(\sqrt{s} = 13\,\text{TeV}\)}",
    journal        = "Eur. Phys. J. C",
    volume         = "76",
    year           = "2016",
    pages          = "292",
    doi            = "10.1140/epjc/s10052-016-4120-y",
    reportNumber   = "CERN-EP-2016-033",
    eprint         = "1603.05598",
    archivePrefix  = "arXiv",
    primaryClass   = "hep-ex",
}

@Article{HIGG-2016-03,
    author         = "{ATLAS Collaboration}",
    title          = "{Search for Higgs boson decays into a pair of light bosons in the \(bb\mu\mu\) final state in \(pp\) collision at \(\sqrt{s} = 13\,\text{TeV}\) with the ATLAS detector}",
    journal        = "Phys. Lett. B",
    volume         = "790",
    year           = "2019",
    pages          = "1",
    doi            = "10.1016/j.physletb.2018.10.073",
    reportNumber   = "CERN-EP-2018-153",
    eprint         = "1807.00539",
    archivePrefix  = "arXiv",
    primaryClass   = "hep-ex",
}

@Article{HIGG-2016-33,
    author         = "{ATLAS Collaboration}",
    title          = "{Measurement of the Higgs boson mass in the \(H \rightarrow ZZ^* \rightarrow 4\ell\) and \(H \rightarrow \gamma\gamma\) channels with \(\sqrt{s} = 13\,\text{TeV}\) \(pp\) collisions using the ATLAS detector}",
    journal        = "Phys. Lett. B",
    volume         = "784",
    year           = "2018",
    pages          = "345",
    doi            = "10.1016/j.physletb.2018.07.050",
    reportNumber   = "CERN-EP-2018-085",
    eprint         = "1806.00242",
    archivePrefix  = "arXiv",
    primaryClass   = "hep-ex",
}

@Article{PERF-2016-04,
    author         = "{ATLAS Collaboration}",
    title          = "{Jet energy scale measurements and their systematic uncertainties in proton--proton collisions at \(\sqrt{s} = 13\,\text{TeV}\) with the ATLAS detector}",
    journal        = "Phys. Rev. D",
    volume         = "96",
    year           = "2017",
    pages          = "072002",
    doi            = "10.1103/PhysRevD.96.072002",
    reportNumber   = "CERN-EP-2017-038",
    eprint         = "1703.09665",
    archivePrefix  = "arXiv",
    primaryClass   = "hep-ex",
}

@Article{PERF-2016-07,
    author         = "{ATLAS Collaboration}",
    title          = "{Performance of missing transverse momentum reconstruction with the ATLAS detector using proton--proton collisions at \(\sqrt{s} = 13\,\text{TeV}\)}",
    journal        = "Eur. Phys. J. C",
    volume         = "78",
    year           = "2018",
    pages          = "903",
    doi            = "10.1140/epjc/s10052-018-6288-9",
    reportNumber   = "CERN-EP-2017-274",
    eprint         = "1802.08168",
    archivePrefix  = "arXiv",
    primaryClass   = "hep-ex",
}

@Article{SUSY-2016-15,
    author         = "{ATLAS Collaboration}",
    title          = "{Search for a scalar partner of the top quark in the jets plus missing transverse momentum final state at \(\sqrt{s} = 13\,\text{TeV}\) with the ATLAS detector}",
    journal        = "JHEP",
    volume         = "12",
    year           = "2017",
    pages          = "085",
    doi            = "10.1007/JHEP12(2017)085",
    reportNumber   = "CERN-EP-2017-162",
    eprint         = "1709.04183",
    archivePrefix  = "arXiv",
    primaryClass   = "hep-ex",
}

@Article{TRIG-2016-01,
    author         = "{ATLAS Collaboration}",
    title          = "{Performance of the ATLAS trigger system in 2015}",
    journal        = "Eur. Phys. J. C",
    volume         = "77",
    year           = "2017",
    pages          = "317",
    doi            = "10.1140/epjc/s10052-017-4852-3",
    reportNumber   = "CERN-EP-2016-241",
    eprint         = "1611.09661",
    archivePrefix  = "arXiv",
    primaryClass   = "hep-ex",
}

@Article{HIGG-2017-05,
    author         = "{ATLAS Collaboration}",
    title          = "{Search for the Higgs boson produced in association with a vector boson and decaying into two spin-zero particles in the \(H \rightarrow aa \rightarrow 4b\) channel in \(pp\) collisions at \(\sqrt{s} = 13\,\text{TeV}\) with the ATLAS detector}",
    journal        = "JHEP",
    volume         = "10",
    year           = "2018",
    pages          = "031",
    doi            = "10.1007/JHEP10(2018)031",
    reportNumber   = "CERN-EP-2018-128",
    eprint         = "1806.07355",
    archivePrefix  = "arXiv",
    primaryClass   = "hep-ex",
}

@Article{HIGG-2017-09,
    author         = "{ATLAS Collaboration}",
    title          = "{Search for Higgs boson decays into pairs of light (pseudo)scalar particles in the \(\gamma\gamma jj\) final state in \(pp\) collisions at \(\sqrt{s} = 13\,\text{TeV}\) with the ATLAS detector}",
    journal        = "Phys. Lett. B",
    volume         = "782",
    year           = "2018",
    pages          = "750",
    doi            = "10.1016/j.physletb.2018.06.011",
    reportNumber   = "CERN-EP-2017-295",
    eprint         = "1803.11145",
    archivePrefix  = "arXiv",
    primaryClass   = "hep-ex",
}

@Article{DAPR-2018-01,
    author         = "{ATLAS Collaboration}",
    title          = "{ATLAS data quality operations and performance for 2015--2018 data-taking}",
    journal        = "JINST",
    volume         = "15",
    year           = "2020",
    pages          = "P04003",
    doi            = "10.1088/1748-0221/15/04/P04003",
    reportNumber   = "CERN-EP-2019-207",
    eprint         = "1911.04632",
    archivePrefix  = "arXiv",
    primaryClass   = "physics.ins-det",
}

@Article{EGAM-2018-01,
    author         = "{ATLAS Collaboration}",
    title          = "{Electron and photon performance measurements with the ATLAS detector using the 2015--2017 LHC proton--proton collision data}",
    journal        = "JINST",
    volume         = "14",
    year           = "2019",
    pages          = "P12006",
    doi            = "10.1088/1748-0221/14/12/P12006",
    reportNumber   = "CERN-EP-2019-145",
    eprint         = "1908.00005",
    archivePrefix  = "arXiv",
    primaryClass   = "hep-ex",
}

@Article{FTAG-2018-01,
    author         = "{ATLAS Collaboration}",
    title          = "{ATLAS \(b\)-jet identification performance and efficiency measurement with \(t\bar{t}\) events in \(pp\) collisions at \(\sqrt{s} = 13\,\text{TeV}\)}",
    journal        = "Eur. Phys. J. C",
    volume         = "79",
    year           = "2019",
    pages          = "970",
    doi            = "10.1140/epjc/s10052-019-7450-8",
    reportNumber   = "CERN-EP-2019-132",
    eprint         = "1907.05120",
    archivePrefix  = "arXiv",
    primaryClass   = "hep-ex",
}

@Article{HDBS-2018-47,
    author         = "{ATLAS Collaboration}",
    title          = "{Search for Higgs boson decays into two new low-mass spin-0 particles in the \(4b\) channel with the ATLAS detector using \(pp\) collisions at \(\sqrt{s} = 13\,\text{TeV}\)}",
    journal        = "Phys. Rev. D",
    volume         = "102",
    year           = "2020",
    pages          = "112006",
    doi            = "10.1103/PhysRevD.102.112006",
    reportNumber   = "CERN-EP-2020-067",
    eprint         = "2005.12236",
    archivePrefix  = "arXiv",
    primaryClass   = "hep-ex",
}

@Article{HIGG-2018-04,
    author         = "{ATLAS Collaboration}",
    title          = "{Observation of \(H \rightarrow b\bar{b}\) decays and \(VH\) production with the ATLAS detector}",
    journal        = "Phys. Lett. B",
    volume         = "786",
    year           = "2018",
    pages          = "59",
    doi            = "10.1016/j.physletb.2018.09.013",
    reportNumber   = "CERN-EP-2018-215",
    eprint         = "1808.08238",
    archivePrefix  = "arXiv",
    primaryClass   = "hep-ex",
}

@Article{HIGG-2018-57,
    author         = "{ATLAS Collaboration}",
    title          = "{Combined measurements of Higgs boson production and decay using up to \(80\,\text{fb}^{-1}\) of proton--proton collision data at \(\sqrt{s} = 13\,\text{TeV}\) collected with the ATLAS experiment}",
    journal        = "Phys. Rev. D",
    volume         = "101",
    year           = "2020",
    pages          = "012002",
    doi            = "10.1103/PhysRevD.101.012002",
    reportNumber   = "CERN-EP-2019-097",
    eprint         = "1909.02845",
    archivePrefix  = "arXiv",
    primaryClass   = "hep-ex",
}

@Article{MUON-2018-03,
    author         = "{ATLAS Collaboration}",
    title          = "{Muon reconstruction and identification efficiency in ATLAS using the full Run~2 \(pp\) collision data set at \(\sqrt{s} = 13\,\text{TeV}\)}",
    journal        = "Eur. Phys. J. C",
    volume         = "81",
    year           = "2021",
    pages          = "578",
    doi            = "10.1140/epjc/s10052-021-09233-2",
    reportNumber   = "CERN-EP-2020-199",
    eprint         = "2012.00578",
    archivePrefix  = "arXiv",
    primaryClass   = "hep-ex",
}

@Article{PIX-2018-001,
    author         = "Abbott, B. and others",
    title          = "{Production and integration of the ATLAS Insertable B-Layer}",
    journal        = "JINST",
    volume         = "13",
    year           = "2018",
    pages          = "T05008",
    doi            = "10.1088/1748-0221/13/05/T05008",
    eprint         = "1803.00844",
    archivePrefix  = "arXiv",
    primaryClass   = "physics.ins-det",
}

@Article{TRIG-2018-01,
    author         = "{ATLAS Collaboration}",
    title          = "{Performance of the ATLAS muon triggers in Run~2}",
    journal        = "JINST",
    volume         = "15",
    year           = "2020",
    pages          = "P09015",
    doi            = "10.1088/1748-0221/15/09/p09015",
    reportNumber   = "CERN-EP-2020-031",
    eprint         = "2004.13447",
    archivePrefix  = "arXiv",
    primaryClass   = "hep-ex",
}

@Article{TRIG-2018-05,
    author         = "{ATLAS Collaboration}",
    title          = "{Performance of electron and photon triggers in ATLAS during LHC Run~2}",
    journal        = "Eur. Phys. J. C",
    volume         = "80",
    year           = "2020",
    pages          = "47",
    doi            = "10.1140/epjc/s10052-019-7500-2",
    reportNumber   = "CERN-EP-2019-169",
    eprint         = "1909.00761",
    archivePrefix  = "arXiv",
    primaryClass   = "hep-ex",
}

@Booklet{ATL-SOFT-PUB-2021-001,
    author         = "{ATLAS Collaboration}",
    title          = "{The ATLAS Collaboration Software and Firmware}",
    howpublished   = "{ATL-SOFT-PUB-2021-001}",
    url            = "https://cds.cern.ch/record/2767187",
    year           = "2021",
}

@Booklet{ATL-SOFT-PUB-2021-003,
    author         = "{ATLAS Collaboration}",
    title          = "{ATLAS Computing Acknowledgements}",
    howpublished   = "{ATL-SOFT-PUB-2021-003}",
    url            = "https://cds.cern.ch/record/2776662",
}

@Report{ATLAS-TDR-2010-19,
    author         = "{ATLAS Collaboration}",
    title          = "{ATLAS Insertable B-Layer Technical Design Report}",
    type           = "ATLAS-TDR-19; CERN-LHCC-2010-013",
    year           = "2010",
    url            = "https://cds.cern.ch/record/1291633",
    related        = "ATLAS-TDR-2010-19-addm",
    relatedstring  = "Addendum:",
}

@Article{CMS-HIG-12-028,
    author         = "{CMS Collaboration}",
    title          = "{Observation of a new boson at a mass of 125 GeV with the CMS experiment at the LHC}",
    journal        = "Phys. Lett. B",
    volume         = "716",
    year           = "2012",
    pages          = "30",
    doi            = "10.1016/j.physletb.2012.08.021",
    reportNumber   = "CERN-PH-EP-2012-220",
    eprint         = "1207.7235",
    archivePrefix  = "arXiv",
    primaryClass   = "hep-ex",
}

@Article{CMS-HIG-14-019,
    author         = "{CMS Collaboration}",
    title          = "{Search for a very light NMSSM Higgs boson produced in decays of the \(125\,\text{GeV}\) scalar boson and decaying into \(\tau\) leptons in \(pp\) collisions at \(\sqrt{s} = 8\,\text{TeV}\)}",
    journal        = "JHEP",
    volume         = "01",
    year           = "2016",
    pages          = "079",
    doi            = "10.1007/JHEP01(2016)079",
    reportNumber   = "CERN-PH-EP-2015-264",
    eprint         = "1510.06534",
    archivePrefix  = "arXiv",
    primaryClass   = "hep-ex",
}

@Article{CMS-HIG-17-024,
    author         = "{CMS Collaboration}",
    title          = "{Search for an exotic decay of the Higgs boson to a pair of light pseudoscalars in the final state with two \(b\) quarks and two \(\tau\) leptons in proton--proton collisions at \(\sqrt{s} = 13\,\text{TeV}\)}",
    journal        = "Phys. Lett. B",
    volume         = "785",
    year           = "2018",
    pages          = "462",
    doi            = "10.1016/j.physletb.2018.08.057",
    reportNumber   = "CERN-EP-2018-089",
    eprint         = "1805.10191",
    archivePrefix  = "arXiv",
    primaryClass   = "hep-ex",
}

@Article{CMS-HIG-17-029,
    author         = "{CMS Collaboration}",
    title          = "{Search for an exotic decay of the Higgs boson to a pair of light pseudoscalars in the final state of two muons and two \(\tau\) leptons in proton--proton collisions at \(\sqrt{s} = 13\,\text{TeV}\)}",
    journal        = "JHEP",
    volume         = "11",
    year           = "2018",
    pages          = "018",
    doi            = "10.1007/JHEP11(2018)018",
    reportNumber   = "CERN-EP-2018-078",
    eprint         = "1805.04865",
    archivePrefix  = "arXiv",
    primaryClass   = "hep-ex",
}

@Article{CMS-HIG-17-031,
    author         = "{CMS Collaboration}",
    title          = "{Combined measurements of Higgs boson couplings in proton--proton collisions at \(\sqrt{s} = 13\,\text{TeV}\)}",
    journal        = "Eur. Phys. J. C",
    volume         = "79",
    year           = "2019",
    pages          = "421",
    doi            = "10.1140/epjc/s10052-019-6909-y",
    reportNumber   = "CERN-EP-2018-263",
    eprint         = "1809.10733",
    archivePrefix  = "arXiv",
    primaryClass   = "hep-ex",
}

@Article{CMS-HIG-18-006,
    author         = "{CMS Collaboration}",
    title          = "{Search for light pseudoscalar boson pairs produced from decays of the \(125\,\text{GeV}\) Higgs boson in final states with two muons and two nearby tracks in \(pp\) collisions at \(\sqrt{s} = 13\,\text{TeV}\)}",
    journal        = "Phys. Lett. B",
    volume         = "800",
    year           = "2020",
    pages          = "135087",
    doi            = "10.1016/j.physletb.2019.135087",
    reportNumber   = "CERN-EP-2019-105",
    eprint         = "1907.07235",
    archivePrefix  = "arXiv",
    primaryClass   = "hep-ex",
}

@Article{CMS-HIG-18-011,
    author         = "{CMS Collaboration}",
    title          = "{Search for an exotic decay of the Higgs boson to a pair of light pseudoscalars in the final state with two muons and two b quarks in \(pp\) collisions at \(13\,\text{TeV}\)}",
    journal        = "Phys. Lett. B",
    volume         = "795",
    year           = "2019",
    pages          = "398",
    doi            = "10.1016/j.physletb.2019.06.021",
    reportNumber   = "CERN-EP-2018-309",
    eprint         = "1812.06359",
    archivePrefix  = "arXiv",
    primaryClass   = "hep-ex",
}

@Article{CMS-HIG-18-024,
    author         = "{CMS Collaboration}",
    title          = "{Search for a light pseudoscalar Higgs boson in the boosted \(\mu\mu\tau\tau\) final state in proton--proton collisions at \(\sqrt{s} = 13\,\text{TeV}\)}",
    journal        = "JHEP",
    volume         = "08",
    year           = "2020",
    pages          = "139",
    doi            = "10.1007/JHEP08(2020)139",
    reportNumber   = "CERN-EP-2020-061",
    eprint         = "2005.08694",
    archivePrefix  = "arXiv",
    primaryClass   = "hep-ex",
}

@Article{CMS-HIG-19-004,
    author         = "{CMS Collaboration}",
    title          = "{A measurement of the Higgs boson mass in the diphoton decay channel}",
    journal        = "Phys. Lett. B",
    volume         = "805",
    year           = "2020",
    pages          = "135425",
    doi            = "10.1016/j.physletb.2020.135425",
    reportNumber   = "CERN-EP-2020-004",
    eprint         = "2002.06398",
    archivePrefix  = "arXiv",
    primaryClass   = "hep-ex",
}

@Booklet{ATLAS-CONF-2015-029,
    author         = "{ATLAS Collaboration}",
    title          = "{Selection of jets produced in \(13~\text{TeV}\) proton--proton collisions with the ATLAS detector}",
    howpublished   = "{ATLAS-CONF-2015-029}",
    url            = "https://cds.cern.ch/record/2037702",
    year           = "2015",
}

@Booklet{ATLAS-CONF-2018-001,
    author         = "{ATLAS Collaboration}",
    title          = "{Measurement of \(b\)-tagging efficiency of \(c\)-jets in \(t\bar{t}\) events using a likelihood approach with the ATLAS detector}",
    howpublished   = "{ATLAS-CONF-2018-001}",
    url            = "https://cds.cern.ch/record/2306649",
    year           = "2018",
}

@Booklet{ATLAS-CONF-2018-006,
    author         = "{ATLAS Collaboration}",
    title          = "{Calibration of light-flavour \(b\)-jet mistagging rates using ATLAS proton--proton collision data at \(\sqrt{s} = 13~\text{TeV}\)}",
    howpublished   = "{ATLAS-CONF-2018-006}",
    url            = "https://cds.cern.ch/record/2314418",
    year           = "2018",
}

@Booklet{ATLAS-CONF-2019-021,
    author         = "{ATLAS Collaboration}",
    title          = "{Luminosity determination in \(pp\) collisions at \(\sqrt{s} = 13\,\text{TeV}\) using the ATLAS detector at the LHC}",
    howpublished   = "{ATLAS-CONF-2019-021}",
    url            = "https://cds.cern.ch/record/2677054",
    year           = "2019",
}

@Article{EvtGen,
    author         = "{D. J. Lange}",
    title          = "{The EvtGen particle decay simulation package}",
    journal        = "Nucl. Instrum. Meth. A",
    volume         = "462",
    year           = "2001",
    pages          = "152",
    doi            = "10.1016/S0168-9002(01)00089-4",
    primaryClass   = "hep-ex",
}
 
\clearpage 
 
\begin{flushleft}
\hypersetup{urlcolor=black}
{\Large The ATLAS Collaboration}

\bigskip

\AtlasOrcid[0000-0002-6665-4934]{G.~Aad}$^\textrm{\scriptsize 99}$,    
\AtlasOrcid[0000-0002-5888-2734]{B.~Abbott}$^\textrm{\scriptsize 124}$,    
\AtlasOrcid[0000-0002-7248-3203]{D.C.~Abbott}$^\textrm{\scriptsize 100}$,    
\AtlasOrcid[0000-0002-2788-3822]{A.~Abed~Abud}$^\textrm{\scriptsize 34}$,    
\AtlasOrcid[0000-0002-1002-1652]{K.~Abeling}$^\textrm{\scriptsize 51}$,    
\AtlasOrcid[0000-0002-2987-4006]{D.K.~Abhayasinghe}$^\textrm{\scriptsize 91}$,    
\AtlasOrcid[0000-0002-8496-9294]{S.H.~Abidi}$^\textrm{\scriptsize 27}$,    
\AtlasOrcid[0000-0001-5329-6640]{H.~Abramowicz}$^\textrm{\scriptsize 157}$,    
\AtlasOrcid[0000-0002-1599-2896]{H.~Abreu}$^\textrm{\scriptsize 156}$,    
\AtlasOrcid[0000-0003-0403-3697]{Y.~Abulaiti}$^\textrm{\scriptsize 5}$
% ...
\bigskip
\\

% ...
$^{5}$High Energy Physics Division, Argonne National Laboratory, Argonne IL; United States of America.\\
% ...
$^{27}$Physics Department, Brookhaven National Laboratory, Upton NY; United States of America.\\
% ...
$^{34}$CERN, Geneva; Switzerland.\\
% ...
$^{51}$II. Physikalisches Institut, Georg-August-Universit\"{a}t G\"ottingen, G\"ottingen; Germany.\\
% ...
$^{91}$Department of Physics, Royal Holloway University of London, Egham; United Kingdom.\\
% ...
$^{99}$CPPM, Aix-Marseille Universit\'e, CNRS/IN2P3, Marseille; France.\\
$^{100}$Department of Physics, University of Massachusetts, Amherst MA; United States of America.\\
% ...
$^{124}$Homer L. Dodge Department of Physics and Astronomy, University of Oklahoma, Norman OK; United States of America.\\
% ...
$^{156}$Department of Physics, Technion, Israel Institute of Technology, Haifa; Israel.\\
$^{157}$Raymond and Beverly Sackler School of Physics and Astronomy, Tel Aviv University, Tel Aviv; Israel.\\
% ...

\end{flushleft}


\end{document}